\documentclass{article}

\usepackage{arxiv}

\usepackage[utf8]{inputenc} 
\usepackage[T1]{fontenc}    
\usepackage{hyperref}       
\usepackage{url}            
\usepackage{booktabs}       
\usepackage{amsfonts}       
\usepackage{nicefrac}       
\usepackage{microtype}      
\usepackage{lipsum}		
\usepackage{graphicx}
\usepackage{doi}

\title{Human-Data Interaction Framework: A Comprehensive Model for a Future Driven by Data and Humans}

\author{ \href{https://orcid.org/0009-0004-2041-7732}{\includegraphics[scale=0.06]{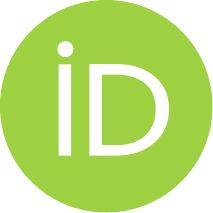}\hspace{1mm}Iván Durango} \\
	  Escuela Superior de Ingeniería Informática\\
	University of Castilla La-Mancha\\
	Albacete, Spain \\
	\texttt{ivan.durango@alu.uclm.es} \\
	\And
	\href{https://orcid.org/0000-0002-6616-8055}{\includegraphics[scale=0.06]{orcid.pdf}\hspace{1mm}José A. Gallud} \\
	Escuela Superior de Ingeniería Informática\\
	University of Castilla La-Mancha\\
	Albacete, Spain \\
	\texttt{jose.gallud@uclm.es} \\
 	\And
	\href{https://orcid.org/0000-0003-1125-9344}{\includegraphics[scale=0.06]{orcid.pdf}\hspace{1mm}Víctor M.R. Penichet} \\
	Escuela Superior de Ingeniería Informática\\
	University of Castilla La-Mancha\\
	Albacete, Spain \\
	\texttt{victor.penichet@uclm.es} \\
}

\renewcommand{\shorttitle}{\textit{arXiv} Template}

\begin{document}
\maketitle

\begin{abstract}
	In an age defined by rapid data expansion, the connection between individuals and their digital footprints has become more intricate. The Human-Data Interaction (HDI) framework has become an essential approach to tackling the challenges and ethical issues associated with data governance and utilization in the modern digital world. This article outlines the fundamental steps required for organizations to seamlessly integrate HDI principles, emphasizing auditing, aligning, formulating considerations, and the need for continuous monitoring and adaptation. Through a thorough audit, organizations can critically assess their current data management practices, trace the data lifecycle from collection to disposal, and evaluate the effectiveness of existing policies, security protocols, and user interfaces. The next step involves aligning these practices with the main HDI principles, such as informed consent, data transparency, user control, algorithm transparency, and ethical data use, to identify gaps that need strategic action. Formulating preliminary considerations includes developing policies and technical solutions to close identified gaps, ensuring that these practices not only meet legal standards, but also promote fairness and accountability in data interactions. The final step, monitoring and adaptation, highlights the need for setting up continuous evaluation mechanisms and being responsive to technological, regulatory, and societal developments, ensuring HDI practices stay up-to-date and effective. Successful implementation of the HDI framework requires multi-disciplinary collaboration, incorporating insights from technology, law, ethics, and user experience design. The article posits that this comprehensive approach is vital for building trust and legitimacy in digital environments, ultimately leading to more ethical, transparent, and user-centric data interactions.
\end{abstract}


\section{Executive Summary}
This article presents a comprehensive field of Human-Data Interaction (HDI), addressing the complex challenges of ethical data management in the digital age. The HDI framework aims to balance the exponential growth of data with the need to respect individual rights, promote transparency, and ensure ethical data practices.
Key components of the framework include:

\begin{itemize}
    \item \textbf{Foundational Principles:} Emphasizing human agency, transparency, fairness, accountability, privacy, and stakeholder engagement.
    \item Structural Components: Outlining data governance frameworks, technological solutions, and human-centric practices necessary for effective HDI implementation.
    \item \textbf{Implementation Roadmap:} Providing a phased approach for organizations to integrate HDI principles, including assessment, policy development, implementation, and continuous adaptation.
    \item \textbf{Ethical and Practical Considerations:} Addressing challenges such as data complexity, bias mitigation, user empowerment, and regulatory compliance.
    \item \textbf{Training and Education:} Stressing the importance of cultivating an HDI-centric culture through comprehensive training programs for employees and education initiatives for users.
    \item \textbf{Monitoring and Adaptation:} Highlighting the need for continuous evaluation and responsiveness to technological, regulatory, and societal shifts.
\end{itemize}

The article argues that adopting HDI principles is not only a moral imperative but also essential for building trust and legitimacy in digital ecosystems. It emphasizes the need for a multidisciplinary approach, combining insights from technology, law, ethics, and user experience design.
Future directions for HDI research and implementation are discussed, including the development of adaptive ethical frameworks, technological innovations, and strategies for global application.

This work contributes to the growing field of ethical data management by providing a structured approach for organizations to navigate the complexities of data interactions in a manner that respects human values and promotes social well-being.

\section{Introduction}
\label{s:intro}
In today's rapidly evolving digital landscape, data has become the lifeblood of our interconnected world. The exponential growth of digital technologies has led to an era in which vast amounts of information are generated, collected, and analyzed at unprecedented scales. From smartphones in our pockets to the expansive networks of smart city infrastructure, data flows endlessly, shaping our daily lives in ways both visible and unseen. This data-driven revolution has led to remarkable advancements in fields ranging from healthcare and scientific research to commerce and public services. Predictive analytics inform critical business decisions, machine learning algorithms power personalized experiences, and big data insights drive policy making at national and global levels. The potential for innovation and progress seems boundless \cite{Tao2018Data}.

However, this data-centric paradigm is not without its challenges. As our reliance on data-driven systems intensifies, we find ourselves grappling with profound ethical, privacy, and security concerns. The Cambridge Analytica scandal laid bare the vulnerabilities of personal data and its potential for manipulation. Algorithmic bias has raised alarming questions about fairness and discrimination in automated decision-making processes. Meanwhile, data breaches continue to erode public trust, highlighting the precarious nature of digital information security.
In addition, the complexity and opacity of many data systems have created a widening gap between those who control these technologies and those affected by them \cite{Wong2019Democratizing}. Individuals often find themselves navigating a digital landscape where their personal information is commodified, their behaviors are predicted and influenced, and their futures are increasingly shaped by algorithms they neither understand nor control.

In this context, a critical juncture is presented. How do we harness the transformative power of data while safeguarding individual rights, fostering transparency, and ensuring ethical practices? How can we build data ecosystems that not only drive innovation, but also empower individuals and promote societal well-being?

These pressing questions underscore the need for a fundamental shift in how we approach data interaction and management. It is against this backdrop that the concept of Human-Data Interaction (HDI) emerges as a crucial paradigm, offering a path towards a more equitable, transparent, and human-centric data future. In response to these pressing challenges, the concept of HDI emerges as a transformative paradigm and represents a fundamental shift in how we approach the relationship between individuals and the data ecosystems that increasingly shape our world. At its core, HDI aims to rebalance the power dynamics in data-driven systems, placing human agency, understanding, and values at the center of data practices \cite{richard_mortier_ab4c0698}.

HDI goes beyond traditional notions of data privacy and security, proposing a holistic framework that encompasses the entire lifecycle of data interactions. It recognizes that individuals are not just subjects of data collection, but active participants in a complex data ecosystem. Using three key principles: legibility, agency and negotiability, HDI seeks to empower individuals with the ability to understand, control and negotiate their data relationships \cite{Rossi2020Transparency}.

Legibility addresses the often opaque nature of data systems, advocating for transparency and comprehensibility in how data are collected, processed, and used. The agency emphasizes the importance of meaningful control that allows individuals to make informed decisions about their data and its implications. Negotiability acknowledges the dynamic nature of data relationships, promoting flexible and context-sensitive approaches to data rights and permissions \cite{Edwards2018Enslaving}.

By integrating these principles, HDI offers a framework for developing more ethical, transparent, and user-centric data practices. It challenges organizations to rethink their data strategies, moving beyond compliance-driven approaches to embrace a more collaborative and responsible ethos. For policymakers, HDI provides a lens through which to craft more nuanced and effective data governance frameworks.

Importantly, HDI is not just a theoretical construct, but a call to action. It requires practical implementation in the technological, organizational, and societal domains. By adopting HDI principles, It is possible to construct data ecosystems that facilitate innovation and advancement while simultaneously upholding human dignity, fostering trust, and contributing to the collective well-being of society..

As we navigate the complexities of our data-driven future, HDI offers a beacon of hope a pathway to harness the power of data while upholding the values that make us human. This article explores the foundations, implications, and practical applications of HDI, charting a course toward a more equitable and empowering data landscape for all.

This article aims to provide a comprehensive exploration and analysis of the Human-Data Interaction (HDI) framework, with the following key objectives and contributions:

\begin{itemize}
    \item \textbf{Conceptual Clarification:}  It provides a comprehensive analysis of the fundamental principles of HDR, offering a clear and accessible explanation of its core tenets legibility, agency, and negotiability. This contributes to a more nuanced understanding of HDI within the broader context of data ethics and governance.
    \item \textbf{Framework Development:} it presents an expanded HDI framework that builds upon existing literature, integrating insights from diverse fields such as computer science, ethics, law, and social sciences. This interdisciplinary approach results in a more robust and applicable model for HDI implementation.
    \item \textbf{Implementation Roadmap:} A significant contribution of this article is the development of a practical, step-by-step roadmap for organizations to adopt and integrate HDI principles. This guide addresses the gap between theoretical understanding and practical application of HDI.
    \item \textbf{Ethical and Practical Considerations}:The ethical implications and practical challenges associated with the implementation of the HDR are analysed in depth. By doing so, It contributes to the ongoing dialogue on responsible data practices and offers strategies to overcome potential obstacles.
    \item Future Directions: It identifies emerging trends and areas for future research in HDR, thereby contributing to the evolution of the field and highlighting opportunities for further academic and practical exploration.
    \item \textbf{Interdisciplinary Synthesis:} By drawing connections between HDI and related fields such as data ethics, privacy law, and user experience design, we contribute to a more holistic understanding of data governance in the digital age.
    \item \textbf{Policy Implications:}The potential impact of the widespread adoption of HDR on data protection policy and regulation is discussed, offering insights that can inform the development of future policy.
\end{itemize}

Through these objectives and contributions, this article aims to advance both the theoretical understanding and practical application of Human-Data Interaction, providing valuable insights for researchers, practitioners, policymakers, and organizations navigating the complexities of ethical data management in the 21st century.

The timeliness and significance of this work cannot be overstated, as HDI has become increasingly crucial in our current technological and societal landscape for several interconnected reasons:

\begin{itemize}
    \item \textbf{Data Ubiquity and Complexity:} The pervasiveness of data collection and processing in everyday life has created what Zuboff (2015) terms "surveillance capitalism" \cite{zuboff2015big}. As our digital footprints expand, HDI provides a framework for managing the complex relationships between individuals and their data. Andrejevic (2014) argues that the sheer volume and variety of data collected about individuals necessitates new approaches to data governance and interaction \cite{andrejevic2014big}.
    \item \textbf{Algorithmic Decision-Making:} With the rise of machine learning and AI, algorithms increasingly influence critical aspects of our lives. Mittelstadt et al. (2016) highlight the ethical challenges posed by algorithmic decision-making, including issues of transparency, fairness, and accountability \cite{mittelstadt2016ethics}. HDI principles, particularly legibility, are crucial for addressing these challenges and ensuring that individuals can understand and challenge decisions made about them.
    \item \textbf{Privacy Concerns:} High-profile data breaches and misuse cases have eroded public trust in data-driven systems. Auxier et al. (2019) report that a majority of Americans feel they have little to no control over their personal data \cite{auxier2019americans}. HDI's emphasis on agency and negotiability offers a pathway to restore trust and give individuals more control over their data.
    \item \textbf{Regulatory Landscape:} The introduction of comprehensive data protection regulations like the GDPR in Europe and the CCPA in California reflects growing societal concerns about data privacy. Cate and Mayer-Schönberger (2013) argue that traditional notice and consent models are insufficient in the big data era \cite{cate2013notice}. HDI provides a more nuanced and flexible approach to data governance that aligns with these evolving regulatory requirements.
    \item \textbf{Digital Divide and Data Literacy:} As data-driven services become more integral to daily life, there's a risk of exacerbating existing inequalities. Ragnedda and Muschert (2013) discuss how differential access to and understanding of digital technologies can reinforce social disparities \cite{ragnedda2013digital}. HDI's focus on legibility and agency can help bridge this gap by making data systems more accessible and understandable to all.
    \item \textbf{Ethical AI and Responsible Innovation:} As AI systems become more advanced, ensuring their ethical development and deployment is paramount. Floridi et al. (2018) propose an ethical framework for AI that aligns closely with HDI principles \cite{floridi2018ai4people}. By centering human values and agency, HDI contributes to the responsible development of AI and other data-driven technologies.
    \item \textbf{Data as a Human Right:} There's growing recognition of data protection and privacy as fundamental human rights. Banisar and Davies (1999) argue for the importance of information privacy in maintaining human dignity and autonomy \cite{banisar1999global}. HDI provides a practical framework for operationalizing these rights in the digital sphere.
    \item \textbf{Personalization vs. Privacy Paradox: }Users often express privacy concerns while simultaneously engaging in privacy-compromising behaviors for perceived benefits. Acquisti et al. (2015) explore this "privacy paradox" \cite{acquisti2015privacy}. HDI's negotiability principle offers a way to navigate this tension by allowing for more nuanced and context-dependent data sharing decisions.
\end{itemize}

As we delve into the complexities of our data-driven world, it becomes increasingly clear that a new paradigm is needed to address the ethical, social, and technical challenges that arise from our interactions with data systems. This necessity forms the foundation of HDI, a concept that seeks to redefine our relationship with data and data-driven technologies. 

To fully grasp the transformative potential of the HDI Framework presented in this article, it is crucial to first understand the fundamental principles and scope of HDI itself. The following section provides a comprehensive definition of Human-Data Interaction, setting the stage for our exploration of its practical applications and far-reaching implications.

\section{Human-Data Interaction Definition}
Human-Data Interaction (HDI) is an emerging field that seeks to place the human at the center of the data flows and processes that characterize our digital world. As defined by Mortier et al. (2014), HDI focuses on the interactions between individuals and data systems, emphasizing the need for these systems to be designed with human values and agency in mind \cite{richard_mortier_ab4c0698}. HDI extends beyond traditional human-computer interaction by specifically addressing the challenges posed by ubiquitous data collection, processing, and use in modern society.
The core principles of HDI, as initially proposed by Mortier et al. and further developed by subsequent researchers, are:

\begin{itemize}
    \item \textbf{Data Legibility:} This principle emphasizes the importance of making data processes and systems understandable to individuals. As Crabtree and Mortier (2015) argue, data legibility involves not just transparency, but also the provision of tools and interfaces that allow people to comprehend how their data is collected, analyzed, and used \cite{andy_crabtree_c8db6d80}. This includes clear communication about data practices and the potential implications of data use.
    \item \textbf{Agency:} The principle of agency focuses on empowering individuals to take action regarding their personal data. This goes beyond mere consent to data collection, advocating meaningful control over data throughout its lifecycle. Luger et al. (2015) highlight that agency in HDI contexts involves giving individuals the ability to make informed decisions about their data, including the right to access, modify, and delete personal information \cite{luger_playing}.
    \item \textbf{Negotiability:} This principle recognizes that the context and meaning of data can change over time and across different situations. As such, HDI advocates for systems that allow for ongoing negotiation of data use and permissions. Coles-Kemp et al. (2018) emphasize that negotiability is crucial for addressing the dynamic nature of privacy preferences and the evolving relationships between individuals and data-driven services \cite{hutton2017}.
 \end{itemize}
These core principles are interconnected and mutually reinforcing. For instance, data legibility is a prerequisite for meaningful agency, as individuals can only make informed decisions about their data if they understand how it is being used. Similarly, negotiability supports both legibility and agency by allowing for adaptable and context-sensitive data relationships.

Recent studies have expanded on these core principles. For example, Victorelli et al. (2020) propose additional dimensions to the HDI framework, including data quality, security, and ethical use \cite{eliane_zambon_victorelli_abe6cbc7}. Their work underscores the evolving nature of HDI and its responsiveness to emerging challenges in the data-driven landscape.

Furthermore, Sailaja et al. (2019) have applied HDI principles to specific domains such as media experiences, demonstrating how these concepts can be operationalized in real-world contexts \cite{sailaja_realising}. Their research highlights the practical implications of HDI and its potential to transform user experiences in data-intensive environments.

By defining and elaborating on these core principles, HDI provides a comprehensive framework for addressing the complex challenges of data ethics, privacy, and user empowerment in the digital age. It offers a foundation for developing more human-centric data practices and technologies that respect individual rights while harnessing the potential of data-driven innovation.

The multifaceted nature of Human-Data Interaction, as defined above, underscores the complexity of the challenges it aims to address. These challenges span across technological, ethical, legal, and social domains, highlighting the limitations of singular, discipline-specific approaches. 

To fully realize the potential of HDI and effectively implement its principles, we must draw upon diverse fields of expertise and foster collaboration across traditional boundaries. This necessity for a holistic perspective leads us to explore the critical importance of interdisciplinary approaches in HDI. By integrating insights from various disciplines, we can develop more comprehensive and effective solutions to the intricate problems posed by our increasingly data-driven world.

\section{Importance of Interdisciplinary Approaches}
The HDI Framework underscores the necessity of employing interdisciplinary approaches to develop comprehensive and effective solutions for the challenges posed by the increasing leverage of personal data \cite{toreini2022}. By integrating perspectives from diverse fields, the framework can harness the complementary expertise and insights required to address the multifaceted nature of human-data interactions.For instance, human-computer interaction can provide valuable insights into user needs, preferences, and behaviors, while data science can offer technical expertise in data manipulation, analysis, and modeling. 

Additionally, fields such as ethics and policy can contribute to the development of ethical frameworks, governance structures, and regulatory mechanisms that ensure the responsible and trustworthy use of personal data \cite{marina_da_bormida_1d0f0927}

The framework also seeks to identify and mitigate the socio-technical challenges that arise from the increasing leverage of personal data in various domains, such as privacy concerns, data transparency and control, accountability issues, and lack of trust.These challenges have been studied and discussed in various disciplines, but effective responses that alleviate them are yet to be fully realized \cite{neelima_sailaja_7c229c50}

The proposed framework takes an interdisciplinary approach, integrating perspectives from fields such as human-computer interaction, data science, ethics, and policy to develop comprehensive and effective solutions \cite{natalia_romero_133dfe31}.By addressing these socio-technical challenges, the framework aims to create data-driven experiences and systems that are aligned with user needs and values, ultimately fostering a more equitable and user-centric data-driven landscape.\cite{martin_maguire_2d9cd775}

The multifaceted nature of Human-Data Interaction necessitates an interdisciplinary approach, drawing insights from diverse fields such as computer science, ethics, law, and social sciences. This interdisciplinary perspective becomes particularly crucial when considering the complex and dynamic nature of the data lifecycle. As data moves through its various stages from collection and processing to storage, utilization, and eventual disposal it intersects with a wide array of disciplines and stakeholders. 

The integration of diverse expertise ensures that each phase of the data lifecycle is approached with a comprehensive understanding of its technical, ethical, and societal implications. For instance, while computer scientists might focus on secure data storage methods, ethicists can provide insights on fair data use, and legal experts can ensure compliance with data protection regulations. By bridging the gap between these disciplines, the HDI framework can more effectively address the challenges that arise at each stage of the data lifecycle, fostering a more holistic and human-centric approach to data management.

\section{The Data Lifecycle}
The Data Lifecycle is a critical component of the HDI Framework as it encapsulates the journey of data from its creation to its eventual disposal. This holistic view ensures that ethical considerations and human-centric practices are applied at every stage of data handling. In the contemporary landscape, data pervades every aspect of our society, influencing diverse domains from personalized service offerings to strategic policy-making on a global scale. The utilization of algorithms, powered by this extensive data, is pivotal in sculpting our daily interactions. They undertake tasks of substantial complexity and consequence, from analyzing medical imagery to curating news feeds, thereby making decisions that carry significant ramifications for society\cite{bruno_lepri_68128501}. Yet, the operationalization of these data-driven frameworks remains an intricate challenge, as they must grapple with the dynamic and relational nature of data itself. 

As data moves through the stages of the lifecycle, from gathering and organizing to analysis and application, it becomes enmeshed with multifaceted social, ethical, and political implications\cite{diana_antonova_87dcc5f8}.The HDI Framework seeks to address these complexities by placing the human at the center of the data lifecycle, ensuring that the design, deployment, and governance of data-driven systems are aligned with user needs,preferences, and values.

Despite the undeniable influence of big data, there exists a peril of neglecting the individual lives encapsulated within these datasets, lives that are invariably affected by the operational systems\cite{siddique_latif_b9e991b5}. While these systems have the potential to yield positive outcomes, the "tyranny of data" can also lead to violations of privacy, information asymmetry, lack of transparency, discrimination, and social exclusion \cite{bruno_lepri_7c9d7b1c}

To mitigate these risks, the proposed HDI Framework advocates for a comprehensive understanding of the data lifecycle, encompassing the processes of collection, storage, processing, and utilization \cite{jessica_rudd_502d802f}By examining each stage of this lifecycle through the lens of human values, needs, and preferences, the framework aims to develop solutions that empower individuals and foster trust in data-driven decision-making \cite{bruno_lepri_7c9d7b1c,neelima_sailaja_7c229c50}.

In reaction to this oversight, the domain of HDI has been proposed, advocating for a data environment that is ethical and empowering. Within this framework, individuals are regarded not as mere subjects of data extraction but as active agents endowed with rights to agency, transparency, and safeguards against data misuse \cite{richard_mortier_ab4c0698}. This discourse posits that the creation of a data ecosystem centered around human interests is not merely a moral obligation but is also essential for sustaining trust and legitimacy in the digital era. 

To this end, the proposed framework delineates the data lifecycle through the lens of human-centric values, encompassing the following key phases:
\begin{enumerate}
\item 
\textbf{Data collection:} Ensuring informed consent and transparency around the purpose and use of collected data \cite{sara_marcucci_d72d29af,natalia_romero_133dfe31}.\item 

\textbf{Data processing:} Enabling users to understand the algorithms and inferences applied to their data, as well as the ability to contest and rectify inaccuracies \cite{sara_marcucci_d72d29af,neelima_sailaja_7c229c50}.\item 

\textbf{Data storage and management: }Providing users with control over the retention and deletion of their personal information, as well as mechanisms for data portability \cite{sara_marcucci_d72d29af,natalia_romero_133dfe31}.\item. 

\textbf{Data utilization:} Empowering users to comprehend how their data is being leveraged and to exercise agency in determining the scope and context of its use \cite{neelima_sailaja_7c229c50}.\item.

\textbf{Data sharing and exchange:} Affording users the right to selectively share their data and to revoke access as needed, fostering a sense of data ownership and control \cite{natalia_romero_133dfe31}.\item

\textbf{Data disposal: }Ensuring the secure and responsible deletion of user data upon request or when it is no longer needed, in alignment with privacy regulations and user preferences \cite{neelima_sailaja_7c229c50}.\item 

\textbf{Data analytics and insights: }Enabling users to understand the inferences and predictions made about them, and to dispute or correct any erroneous conclusions \cite{natalia_romero_133dfe31,neelima_sailaja_7c229c50}.\item 

\textbf{Ongoing monitoring and evaluation:} Establishing mechanisms for continuous assessment and improvement of the data ecosystem, addressing evolving user needs and emerging technological advancements \cite{natalia_romero_133dfe31}.
\end{enumerate}

The HDI Framework proposes a comprehensive, human-centric approach to the data lifecycle, aiming to instill individual empowerment and collective accountability in our data-driven landscape \cite{marina_da_bormida_1d0f0927}. This paradigm shift goes beyond mere technical solutions, outlining a structured approach for embedding ethical principles throughout the entire data journey.

Central to this framework is the development of stringent policies upholding informed consent and data sovereignty, the integration of user-empowerment tools, and the promotion of data ethics education \cite{sara_marcucci_d72d29af}. Despite the complexities of modern data systems and the risk of algorithmic bias \cite{sorelle_a_friedler__carlos__scheidegger__suresh__venkatasubramanian_undefined_6b9b97b9}, the adoption of HDI principles offers profound benefits, positioning organizations to utilize data responsibly, foster innovation, and contribute to societal well-being.

The HDI Framework represents a transformative shift, placing individuals at the core of shaping our data-driven future. By prioritizing user agency, transparency, and ethics, it lays the groundwork for a more equitable and empowering data ecosystem \cite{matteo_turilli_24f7a58d}.

Human engagement is crucial throughout the data lifecycle to ensure ethical, equitable, and socially beneficial data practices \cite{woocheol_kim_e552fe9a}. This involves empowering individuals with agency and choice in data collection, processing, and utilization, while fostering transparency and accountability in data-driven systems.
At the collection stage, individuals should receive comprehensive information about the purpose, scope, and intended use of their data \cite{donghao_zhou_8dc5bba2}. This transparency enables informed consent and allows for selective data sharing based on personal preferences.
During processing and analysis, individuals must be empowered to understand the algorithms and processes applied to their data, including potential implications. This legibility allows users to contest inaccuracies and understand how their data is being used \cite{neelima_sailaja_7c229c50}.

In storage and management, individuals should have control over data retention and deletion, with data portability features allowing seamless transfer between service providers \cite{natalia_romero_133dfe31}.The utilization and sharing phases require balancing individual privacy with potential societal benefits from data-driven insights. Individuals should be able to selectively share data while maintaining control over access and scope.
By embedding human agency, transparency, and control throughout the data lifecycle, the HDI Framework fosters a more equitable and empowering data ecosystem \cite{richard_mortier_ab4c0698}. 
This approach safeguards individual rights while incentivizing responsible organizational data practices.
Crucially, the application of user data in decision-making must include mechanisms for user oversight and the ability to challenge decisions, empowering individuals to shape how their information is used and fostering accountability and transparency \cite{maxwell_zostant_786f4300}. 

Incorporating this lifecycle approach into the HDI Framework ensures consistent application of human-centric principles across all data-related activities, enhancing both ethical practices and overall data management efficiency.

While the Data Lifecycle provides a comprehensive framework for managing data ethically throughout its existence, implementing such a structure requires careful planning and strategic considerations. As organizations prepare to embed HDI principles across the data lifecycle, they must first address several preliminary considerations. 

These foundational steps ensure that the organization is well-positioned to successfully integrate HDI practices into every stage of the data journey. By thoughtfully examining these preliminary considerations, organizations can create a solid groundwork for implementing a human-centric approach to data management that aligns with ethical standards, regulatory requirements, and stakeholder expectations. 

The following section outlines these crucial preliminary considerations, which serve as a bridge between understanding the theoretical importance of the data lifecycle and practically implementing HDI principles within an organization's unique context.

\section{Preliminary Considerations}

Before implementing a Human-Data Interaction (HDI) framework, organizations must carefully evaluate and prepare their existing structures, processes, and culture. These preliminary considerations serve as a crucial foundation, ensuring that the subsequent implementation of HDI principles is well-informed, strategically sound, and aligned with both organizational values and ethical data practices. By thoroughly addressing these initial aspects, organizations can pave the way for a more successful and sustainable integration of HDI principles into their core operations. The following subsections outline key areas that demand attention in this preparatory phase. This stage involves laying the groundwork for implementing the insights gleaned from alignment analysis, focusing on ethical, technical, and operational considerations to improve data practices. This proactive approach ensures that the subsequent steps taken are informed, deliberate, and aligned with both organizational values and HDI principles.

\subsection{Ethical Considerations}
At the forefront of preliminary considerations is the emphasis on ethics. Organizations must deliberate on the ethical implications of their data practices, particularly in terms of respecting user autonomy, ensuring privacy, and preventing harm\cite{david_j__hand_687e598b}.This involves considering how data is collected, used, and shared, ensuring that these practices are not only legally compliant but also ethically sound. The work of Shilton\cite{katie_shilton_287b89c9} underscores the importance of embedding ethical considerations into the fabric of organizational culture and decision-making processes, ensuring that data practices reflect a commitment to respect, fairness, and responsibility toward individuals whose data are being handled. 

To this end, organizations should establish a cross-functional ethics review board or advisory panel, tasked with evaluating data-driven initiatives through the lens of HDI principles.

\subsection{Technical Considerations}
Technical considerations form the backbone of effective HDI implementation. Organizations must assess the capabilities and limitations of their existing technological infrastructure, identify areas for improvement, and ensure that data practices are underpinned by robust, user-centric systems.

This may involve enhancing data management and governance frameworks, implementing user-friendly data access and control interfaces, and deploying advanced analytics and visualization tools that enhance data legibility and transparency. 

Furthermore, organizations should explore the potential of emerging technologies, such as blockchain and secure multi-party computation, to bolster data security, provenance, and user control \cite{linh_t__nguyen_lam_duc_nguyen_thong_hoang_h__m__n__dilum_bandara_qin_wang_qinghua_lu_xinhua_xu_zhu_petar_popovski_shiping_chen_undefined_934a84c2}

Organizations must consider investments in technologies that enable more granular consent mechanisms, robust encryption, and secure data storage and transmission protocols\cite{siani_pearson_ce18a5ef}. In addition, the development or adoption of tools for bias detection and mitigation in algorithms, as well as those that enhance the explainability of machine learning models, is critical. These technical upgrades and innovations are essential to operationalize the HDI principles in a way that is both effective and sustainable.

\subsection{Operational Considerations}
Operational considerations encompass the processes and policies that govern data practices within the organization. This includes re-visiting data governance frameworks to explicitly incorporate HDI principles, ensuring that data collection, storage, use, sharing, and disposal policies are aligned with ethical standards and technical capabilities\cite{zeeshan_khan_8f0c5da8}. The establishment of clear, transparent communication channels for stakeholders, including data subjects, is crucial for fostering trust and accountability. 

Operational considerations also involve the training and engagement of employees across the organization, from leadership to frontline staff, in understanding and championing HDI principles. The work of Burr et al.\cite{christopher_burr_33521b24} highlights the importance of integrating legal, technical, and policy expertise in crafting and implementing data practices that are not only compliant but also ethically and socially responsible.

\subsection{Stakeholder Engagement}
Engaging with stakeholders, including users, regulatory bodies, and civil society organizations, is a critical component of preliminary considerations. This engagement ensures that the organization's efforts to implement HDI principles are responsive to the needs and concerns of those affected by its data practices\cite{neelima_sailaja_361ac9d8}. Through workshops, focus groups, and other participatory methods, organizations can gain valuable insights into user expectations, preferences, and concerns regarding data management.

User participation, in particular, is essential to understand the unique needs, preferences, and concerns of individuals whose data are being collected and used. This may involve conducting user research, collecting feedback on data practices, and establishing mechanisms for ongoing dialogue and collaboration\cite{jimmy_moore_e7ccf9c9}.Engagement with regulatory bodies and civil society organizations can provide valuable insights into emerging trends, best practices, and potential pitfalls in HDI implementation, thus guiding the organization's efforts and ensuring compliance with relevant laws and regulations.

Stakeholder feedback can provide valuable insights into potential ethical dilemmas, privacy concerns, and areas for improvement in transparency and accountability. Furthermore, this engagement fosters a collaborative approach to data governance, improving the legitimacy and effectiveness of the organization's data practices.

\subsection{Alignment Analysis}
The alignment analysis phase involves a comprehensive assessment of the organization's existing data practices against the principles of HDI. This analysis serves as a foundational step in identifying areas of alignment, gaps, and opportunities for improvement, forming the subsequent implementation of HDI-centric initiatives \cite{georges_nguefack_tsague_552e7b48}.

The analysis should consider the organization's data ecosystem holistically, examining factors such as data collection, storage, usage, sharing, and disposal practices. Additionally, the analysis should assess the organization's technological capabilities, governance frameworks, and stakeholder engagement mechanisms to identify areas of alignment and opportunities for enhancement.

Mapping an organization's data practices to HDI principles begins with a comprehensive review of current policies, procedures, and technical systems related to data management. This review should consider factors such as the organization's data collection methods, storage and processing practices, data sharing and disclosure protocols, and the mechanisms in place for user control and transparency. By conducting a thorough alignment analysis, organizations can gain a clear understanding of their current state and develop a roadmap to effectively implement the HDI principles.

\subsection{Audit Existing Practices}
The initiation of a HDI framework within an organization necessitates a foundational step: a meticulous audit of existing data handling practices. This critical examination serves not merely as a procedural checkpoint but as a profound inquiry into the organization's data lifecycle management, from inception to obsolescence.

By undertaking this audit, an organization can delineate a comprehensive baseline from which to enhance and refine its data stewardship, guided by the tenets of HDI.
\begin{itemize}
\item 

\textbf{Comprehensive Data Mapping: }The audit commences with an exhaustive mapping of the data lifecycle, illustrating the flow of data from its collection points through to its final disposition. This visual documentation encompasses identifying all data ingress points, such as sensors, web interfaces, and third-party integrations, and tracing the data's journey across storage, processing, internal sharing, and external disclosures. Humayun et al.\cite{mamoona_humayun_9e268e7c} underscore the importance of understanding these data flows, highlighting how such mapping can unveil the intricacies and potential vulnerabilities within an organization's data ecosystem.\item 

\textbf{Policies and Procedures Analysis:} A thorough review of the existing data governance frameworks is imperative. This includes scrutinizing privacy policies, data handling procedures, and terms of service to ascertain their alignment with HDI principles such as informed consent, data minimization, and user rights\cite{nili_steinfeld_651aabd0}. The clarity, accessibility, and comprehensiveness of these documents are evaluated to ensure they serve not only as legal compliance mechanisms but also as transparent guides for users.\item 

\textbf{Security Measures Evaluation: }Assessing the robustness of data security measures forms a critical part of the audit. Organizations must evaluate their encryption protocols, access control mechanisms, and incident response strategies to identify vulnerabilities\cite{siani_pearson_ce18a5ef}. This assessment helps in pinpointing areas where enhancements are necessary to fortify the organization against data breaches and unauthorized access.\item 

\textbf{Data-Driven Decision-Making Scrutiny:} The use of data in decision-making processes, particularly when involving algorithms or machine learning models, requires rigorous examination. The audit should investigate these models for biases and their explainability, ensuring that decision-making processes are transparent and equitable\cite{sorelle_a_friedler_9b8f9f6f}. This scrutiny aims to guarantee that human oversight is effectively integrated, allowing for intervention and accountability.\item 

\textbf{User Interface Inspection:} Finally, the interfaces through which users interact with the organization's data systems warrant careful evaluation. This involves assessing the clarity and efficacy of consent mechanisms, the ease with which users can access or amend their data, and the transparency surrounding the usage of their data \cite{wendy_lucas_91b77e43}
\end{itemize}

This auditing process, necessitating a collaborative effort among diverse organizational teams, embodies a multidisciplinary approach. By integrating insights from technical staff, legal experts, and user experience designers, organizations can gain a holistic understanding of their current data handling practices.\cite{qian_wang_16794540}

Such a comprehensive audit not only illuminates existing strengths and weaknesses but also paves the way for the development of a nuanced HDI framework. Through targeted interventions, grounded in the insights garnered from the audit, organizations can align their data practices with core values and goals, fostering a culture of ethical data stewardship and enhancing trust among users.

Having examined these crucial preliminary considerations, organizations can lay a solid foundation for implementing a comprehensive HDI framework. These initial steps provide the necessary context and groundwork for a more detailed and structured approach to integrating HDI principles throughout an organization's data practices. 

Building upon this preparatory phase, we now turn to a detailed exploration of the HDI Framework Schema. This schema offers a comprehensive, process-based model that operationalizes HDI principles, providing organizations with a practical guide to embed ethical data practices into their core operations. By progressing from preliminary considerations to this structured framework, organizations can systematically transform their approach to data management, ensuring that human values and ethical considerations are central to all data-related activities.

\section{HDI Framework Schema}
The proposed HDI framework is a comprehensive process-based model designed to address the complex challenges of data management and utilization. Building upon the foundational, this framework expands and operationalizes the HDI principles to provide a practical guide for organizations, policymakers, and researchers.

The framework is structured around nine key components, emphasizing its process-oriented nature:

\begin{enumerate}
    \item \underline{\textbf{Foundational Principles}}
    \begin{itemize}
    \item \textbf{Agency:} Empowering individuals with control over their data.
    \item \textbf{Legibility:} Making data processes transparent and understandable.
    \item \textbf{Equity} Ensuring fair and unbiased data practices.
    \item \textbf{Accountability:} Clearly defining responsibility for data use.
    \item \textbf{Privacy and Data Protection:} Safeguarding personal information.
    \item \textbf{Participation and Engagement:} Meaningful stakeholder involvement throughout the data lifecycle, from design to deployment.
    \item \textbf{Negotiability:} Allowing flexible, context-based data permissions.
    \item \textbf{Human-Centricity:} Prioritizing human needs in data systems.
    \item \textbf{Consent:} Ensuring informed and meaningful consent for data collection and use.
    \end{itemize}
    
    \item \underline{\textbf{Structural Components}}
    \begin{itemize}
    \item \textbf{Data Governance}: Establishing policies for ethical data management.
    \item \textbf{Technological Infrastructure:} Implementing tools supporting HDI principles.
    \item \textbf{Organizational Culture:} Fostering an ethical data mindset.
    \item \textbf{Stakeholder Engagement:} Involving users in data decisions.
    \item \textbf{Ethical Review Processes:} Continuously assessing data practices.
    \item \textbf{Data Ethics Board:} Establishing a dedicated group to oversee ethical data practices.
    \item \textbf{Data Quality Management:} Ensuring data accuracy and relevance.
    \end{itemize}

    \item \underline{\textbf{Ethical Considerations}}
    \begin{itemize}
        \item \textbf{Respect for Individual Autonomy:} Honoring individual choices.
        \item \textbf{Equity and Justice:} Promoting fairness in data practices.
        \item \textbf{Transparency:} Ensuring openness in data processes.
        \item \textbf{Responsible Innovation:} Ethically advancing data technologies.
        \item \textbf{Environmental Impact:} Considering sustainability in data management.
        \item \textbf{Bias Mitigation:} Addressing prejudices in data and algorithms.
        \item \textbf{Ethical AI Use:} Aligning AI with HDI principles.
        \item \textbf{Cultural Sensitivity:} Considering diverse cultural perspectives in data practices.
    \end{itemize}
    
    \item \underline{\textbf{Practical Considerations}}
    \begin{itemize}
        \item \textbf{System Integration:} Incorporating HDI into existing infrastructures.
        \item \textbf{ Technological Feasibility:} Assessing implementability of HDI solutions.
        \item \textbf{Scalability:} Ensuring HDI practices can grow with the organization.
        \item \textbf{Regulatory Compliance:} Meeting legal data requirements.
       \item \textbf{ Organizational Capacity:} Evaluating readiness for HDI adoption.
        \item \textbf{Education and Training:} Preparing staff for HDI implementation.
        \item \textbf{Cost-Benefit Analysis:} Assessing financial implications of HDI.
        \item \textbf{Risk Management:} Identifying and mitigating data-related risks.
        \item \textbf{Interoperability:} Ensuring HDI practices work across different systems and platforms.
    \end{itemize}

    \item \underline{\textbf{Implementation Roadmap }}
    \begin{itemize}
        \item \textbf{Assessment: } Evaluating current data practices against HDI principles. 
        \item \textbf{Design: } Developing HDI-aligned systems and processes.
        \item \textbf{Integration: } Implementing HDI practices across the organization.
        \item \textbf{Evaluation:} Continuous monitoring and improvement of HDI initiatives.
        \item \textbf{Adaptation:} Updating practices for emerging challenges.
        \item \textbf{Pilot Testing:} Implementing HDI practices in small-scale projects before full deployment.
    \end{itemize}
    
    \item \underline{\textbf{Enhancing HDI Culture}}
    \begin{itemize}
        \item \textbf{Employee Training Programs}: Educating staff on HDI principles.
        \item \textbf{Stakeholder Education:} Broadening the scope to include all affected parties, not just direct users.
        \item \textbf{Continuous Learning and Adaptation:} Updating training based on new trends.
        \item \textbf{Cross-functional Collaboration:} Fostering interdepartmental HDI understanding.
    \end{itemize}
    
    \item \underline{\textbf{Data Lifecycle Management}}
    \begin{itemize}
        \item \textbf{Collection:} Gathering data ethically and transparently.
        \item \textbf{Processing:} Analyzing data fairly and accountably.
        \item \textbf{Storage:} Retaining data securely and responsibly.
        \item \textbf{Utilization:} Using data ethically in decision-making.
        \item \textbf{Sharing:} Exchanging data with consent and control.
        \item \textbf{Disposal:} Deleting data properly and timely.
    \end{itemize}

    \item \underline{\textbf{Monitoring and Adaptation}}
    \begin{itemize}
        \item \textbf{Continuous Assessment:} Regularly evaluating HDI practices.
        \item \textbf{Feedback Mechanisms:} Gathering stakeholder input on data practices.
        \item \textbf{Adaptive Governance:} Evolving policies with changing needs.
        \item \textbf{Emerging Technology Integration:} Incorporating new HDI-enhancing tools.
        \item \textbf{Performance Metrics:} Tracking HDI-specific key performance indicators.
       \item \textbf{ External Audits: Conducting independent assessments of HDI practices.}
    \end{itemize}
        
    \item \underline{\textbf{Complex and dynamic considerations}}
    \begin{itemize}
        \item \textbf{Complexity:} Acknowledging the intricate nature of data systems and the need for nuanced, context-specific HDI implementations.
        \item \textbf{Evolving Technologies:} Addressing the continual emergence of new data technologies and their implications for HDI.
        \item \textbf{Regulatory Landscape:} Keeping pace with changing data protection laws and regulations across different jurisdictions.
        \item \textbf{Societal Shifts:} Recognizing changing public attitudes towards data privacy and use.
        \item \textbf{Ethical Dilemmas:} Preparing for unforeseen ethical challenges that may arise with new data uses.
        \item \textbf{Cross-border Data Flows:} Managing HDI principles in the context of global data transfer and varying international standards.
        \item \textbf{Balancing Innovation and Ethics:} Navigating the tension between rapid technological advancement and ethical considerations.
        \item \textbf{Adaptive Strategies:} Developing flexible approaches that can evolve with changing data ecosystems.
    \end{itemize}
    
\end{enumerate}

This process-based framework is designed to be iterative and adaptable across various sectors and organizational sizes. It recognizes the dynamic nature of the data landscape, as highlighted by \cite{cate2013notice}, and provides flexibility to address emerging challenges.

The framework incorporates insights from recent research on algorithmic accountability \cite{mittelstadt2016ethics} and privacy economics \cite{Acquisti2015}, ensuring a holistic approach to data ethics and management. It emphasizes the importance of continuous learning and adaptation, acknowledging that HDI implementation is an ongoing process rather than a one-time initiative.

By integrating these components into a cohesive, process-oriented approach, the proposed HDI framework offers a roadmap for organizations to navigate the complexities of data ethics, privacy, and user empowerment. It aims to foster a data ecosystem that not only drives innovation and efficiency but also respects human values and promotes societal well-being.

This framework represents a significant step towards realizing the vision of a human-centric data society, where technological advancements and ethical considerations evolve in tandem, as advocated \cite{floridi2018ai4people} in their work on AI ethics. Its process-based nature ensures that organizations can continually refine and improve their HDI practices in response to new challenges and opportunities in the ever-evolving data landscape.

\section{Foundational Principles}
The HDI Framework presents a comprehensive and multidimensional approach to managing and utilizing data in a manner that prioritizes ethical considerations, upholds fundamental human rights, and promotes the overall well-being of society. This holistic framework guides organizations in cultivating a data ecosystem that deeply respects individual autonomy, empowering users with meaningful control over their personal information, including its collection, usage, sharing, and deletion.\cite{Sujata_e7534125}

The framework emphasizes the critical importance of transparency, requiring clear and comprehensible explanations regarding data handling practices, algorithmic decision-making processes, and the security protocols in place to safeguard data integrity\cite{bartha_maria_knoppers_2f6be343}. Ensuring fairness and non-discrimination is a core tenet, demanding proactive efforts to identify and mitigate biases within datasets and algorithmic systems, thereby promoting equitable outcomes for all individuals. Accountability is woven throughout the framework, with mechanisms established to address instances of data misuse, facilitate redress for affected parties, and demonstrate organizational responsibility. 

Additionally, the framework underscores the value of meaningful stakeholder engagement, advocating for the inclusion of diverse perspectives and the co-creation of ethical data practices with users and affected communities. By outlining these guiding principles, structural components, implementation strategies, and multifaceted considerations\cite{giselle_corbie_smith_a8034ceb}, the HDI Framework provides a comprehensive blueprint for organizations to embed ethical practice and user-centric values at the very core of their data management and data-driven decision-making processes.

\subsection{Agency}
Human agency and autonomy are central to the HDI framework, emphasizing the empowerment of individuals in their interactions with data-driven systems. This principle advocates for giving individuals control over their data, including its collection, use, sharing, and deletion. It underscores the importance of enabling people to make informed decisions about their data, reflecting a commitment to respecting personal autonomy and privacy \cite{privacy_by_design_the_7_foundational_principles_4004e322}

Individuals have the right to exercise control over their personal data, including the right to access, modify, and delete their information, as well as to make informed decisions about data collection and use. Respecting autonomy acknowledges individuals as active agents with the capacity to make choices about their data. This principle is fundamental to ensuring that data practices empower rather than undermine user agency.\cite{lindsay_stirton_46af6e46}

\subsection{Legibility}
Transparency and explainability within HDI demand clear, understandable information about data practices, the workings of algorithms, and the security measures in place to protect data\cite{kathleen_creel_94e92dc7}. This principle is crucial for building trust and accountability, allowing users to grasp how their data is processed and how decisions that affect them are made. Explainable AI (XAI) systems play a pivotal role in demystifying algorithmic processes, making technology accessible and comprehensible to non-experts (Ribeiro, Singh, And Guestrin, 2016).\cite{stefaan_verhulst_565565b1}.

Maintain openness about data practices, including the collection, use, and sharing of data, and implement mechanisms for accountability in decision-making processes. Transparency and accountability are vital for building trust, facilitating oversight, and enabling individuals to understand and challenge decisions made with or about their data \cite{marina_da_bormida_1d0f0927,stefaan_g_verhulst_undefined_6313dfb3}

\subsection{Equity} 
Fairness and non-discrimination call for proactive efforts to detect and mitigate biases in datasets and algorithms, ensuring equitable outcomes for all users\cite{sandra_wachter_a92be820,Schwartz_b490bbb2}. This principle recognizes the potential for data-driven systems to perpetuate or even exacerbate existing inequalities. Research in algorithmic fairness seeks to develop methodologies and tools to identify and correct biases, promoting fairness in automated decision-making processes \cite{jie_xu_fef7d595}

Ensure equitable access to the benefits of data technologies and protect against discriminatory practices and biases that may arise from data collection, analysis, and application. Addressing justice and equity is crucial for mitigating the risk of exacerbating existing inequalities and ensuring that the benefits of data technologies are shared across all sectors of society \cite{david_leslie_11e3ae41}

\subsection{Accountability}
Accountability and redress in the HDI framework ensure that mechanisms are in place to address any misuse of data, provide remedies to affected individuals, and demonstrate organizational responsibility\cite{anthony_vance_409afbec}. This principle involves establishing clear policies and procedures for data governance, including how data breaches and privacy violations are handled. It also encompasses the need for regulatory compliance and the ability to audit and review data practices to uphold ethical standards \cite{charna_r__parkey_382d16a9}

Data practices should aim to benefit individuals and communities while minimizing harm and preventing misuse of data that could lead to adverse outcomes. The dual obligation to do good and avoid harm is a cornerstone of ethical practice, emphasizing the responsibility of data handlers to consider the broader impact of their actions on society.\cite{nicki_tiffin_5528c478}

\subsection{Privacy and Data Protection}
Privacy and data protection are foundational to HDI, safeguarding individuals' rights to control their personal information and ensuring that data is handled securely and respectfully\cite{soheil_human_13bdc9ca}. This principle is aligned with global privacy regulations, such as the General Data Protection Regulation (GDPR) in the European Union, which set standards for data collection, processing, and storage. Implementing privacy-by-design approaches and privacy-enhancing technologies (PETs) are practical steps towards operationalizing this principle.

Uphold individuals' rights to privacy and data protection, implementing secure and responsible data handling practices to safeguard personal information from unauthorized access and exploitation.

Privacy is essential for personal security, freedom of expression, and protection against surveillance and control, making it a fundamental aspect of ethical data practices.

\subsection{Participation and Engagement}
Participation and engagement advocate for the inclusion of stakeholders, including data subjects, in the design and governance of data-driven systems\cite{vincent_luyet_b9708be0}. This principle emphasizes the value of diverse perspectives and the importance of co-creating ethical data practices with users and affected communities. It supports the idea of democratizing data, where individuals and communities have a say in how data about them is used and for what purposes.

Foster inclusive and participatory approaches that engage stakeholders, particularly those most affected by data practices, in the governance, design, and implementation of data technologies. Participation and empowerment ensure that diverse perspectives are considered, supporting democratic values and enhancing the relevance and acceptability of data initiatives. Engaging with users and impacted communities helps to ensure that data practices are responsive to their needs and concerns, promoting transparency, trust, and shared responsibility \cite{rachel_charlotte_smith_e160fda8}

Implementing these foundational principles requires a multidisciplinary approach, combining technical solutions with ethical, legal, and social insights. The HDI framework aims to foster a data ecosystem that respects individual rights and promotes social good, ensuring that technology serves humanity in equitable and responsible ways.

\subsection{Negotiability}
The principle of negotiability in HDI recognizes that the context and meaning of data can change over time and across different situations. It emphasizes the need for flexible, dynamic approaches to data permissions and use. Negotiability allows for ongoing adjustments to data sharing preferences, acknowledging that user priorities and circumstances may evolve \cite{richard_mortier_ab4c0698}.
Implementing negotiability involves:

\begin{itemize}
\item Developing mechanisms for users to revise their data sharing preferences over time.
\item Creating systems that allow for context-dependent data permissions.
\item Establishing protocols for renegotiating data use agreements as new uses or insights emerge.
\end{itemize}

Mortier et al. argue that negotiability is crucial for addressing the dynamic nature of privacy preferences and the evolving relationships between individuals and data-driven services \cite{richard_mortier_ab4c0698}. This principle aligns with the concept of "contextual integrity" proposed by Nissenbaum, which posits that the appropriateness of information flows depends on context-specific norms \cite{nissenbaum2004privacy}.

\subsection{Human-Centricity}
Human-centricity places human needs, values, and experiences at the core of data system design and operation. This principle ensures that technological advancements serve human interests rather than subordinating human agency to algorithmic imperatives \cite{shneiderman2020human}.

Key aspects of human-centricity in HDI include:

\begin{itemize}
\item Prioritizing user experience and accessibility in data interfaces.
\item Considering the psychological and social impacts of data practices on individuals and communities.
\item Ensuring that data-driven systems augment rather than replace human decision-making capacities.
\end{itemize}

Shneiderman advocates for a human-centered approach to AI and data systems, emphasizing the importance of keeping humans "in the loop" and designing systems that enhance human capabilities \cite{shneiderman2020human}. This aligns with the broader movement towards "human-centered AI" that seeks to develop technologies that are beneficial, comprehensible, and controllable by humans \cite{li2018human}.

\subsection{Consent}
Informed consent is a foundational principle in HDI, rooted in respect for individual autonomy and self-determination. It requires that individuals be provided with clear, understandable information about data collection, use, and potential implications before agreeing to share their data \cite{solove2013introduction}.
Effective implementation of consent in HDI frameworks involves:

\begin{itemize}
\item Developing clear, accessible consent mechanisms that go beyond mere legal compliance.
\item Providing granular options for consent, allowing users to selectively agree to specific data uses.
\item Ensuring that consent is ongoing and revocable, with easy-to-use tools for withdrawing or modifying consent.
\item Educating users about the implications of their consent decisions.
\end{itemize}

Recent research has highlighted the limitations of traditional notice and consent models in the big data era, leading to proposals for more dynamic and user-centric approaches to consent \cite{custers2016dynamic}. These new models aim to address the challenges of informed consent in complex, rapidly evolving data ecosystems.

The foundational principles of Human-Data Interaction (HDI) outlined above—agency, legibility, equity, accountability, privacy, participation, negotiability, human-centricity, and consent—provide the ethical and conceptual framework upon which robust HDI practices are built. However, translating these principles into practical, operational realities within organizations requires a set of concrete structural components. 

These components serve as the architectural elements that enable the realization of HDI principles in real-world data ecosystems. As we transition from the theoretical underpinnings to the practical implementation of HDI, it becomes crucial to examine how these foundational principles can be embedded into organizational structures, technological systems, and governance frameworks. 

The following section on Structural Components explores the tangible mechanisms and processes that organizations can adopt to operationalize HDI principles, ensuring that ethical considerations are not merely aspirational but are deeply integrated into the fabric of data management and utilization practices. By bridging the conceptual with the practical, we can begin to envision how HDI can transform from a set of ideals into a lived reality within complex organizational environments.

\section{Structural Components}
The structural components of the HDI framework provide a tangible architecture for embedding ethical principles into the practical operations of data-driven systems. These components are critical for translating ethical commitments into actionable strategies and processes that can be systematically applied within organizations. This section outlines the key structural elements essential for establishing a robust HDI practice, drawing on interdisciplinary research to support their implementation.

\subsection{Data Governance}
A comprehensive data governance is essential for ensuring that data management practices align with ethical principles and regulatory requirements. This framework encompasses policies, standards, and procedures that guide how data is collected, stored, accessed, and used within an organization.
\begin{itemize}
\item 

\textbf{Policies and Protocols}: Develop and implement clear policies that cover informed consent, data minimization, purpose limitation, data quality, retention, and deletion. These policies should also address security measures and incident response plans to protect data integrity and confidentiality.\cite{nicki_tiffin_5528c478}\item 

\textbf{Audit and Compliance:} Establish regular audit processes to evaluate the alignment of data practices with HDI principles, legal standards, and evolving societal expectations\cite{johan_frishammar_bf19fa7e}. Compliance checks should ensure adherence to internal policies and external regulations, facilitating accountability and continuous improvement \cite{raymond_j__burby_cee9fc1c}
\end{itemize}
\begin{itemize}
\item 

\textbf{Roles and Responsibilities:} Define clear roles and responsibilities for data stewardship, including data owners, data stewards, and data governance committees. These roles should be empowered to make decisions, enforce policies, and address issues that arise in the course of data management.\cite{doris_hooi_ten_wong_e0ef9064}
\end{itemize}

\subsection{Technological Infrastructure}
Technological tools and systems play a crucial role in facilitating ethical data practices. Using technology, organizations can improve transparency, protect privacy, and empower users.
\begin{itemize}
\item 

\textbf{Consent Management Platforms (CMPs):} Deploy CMPs to streamline the process of obtaining and managing user consent for data collection and processing, ensuring that consent is informed, specific, and revocable \cite{eve_maler_ed33c9f3}\item 

\textbf{Data Visualization Tools:} Utilize data visualization tools to make data practices more transparent and comprehensible to non-expert stakeholders, aiding in the communication of complex information.\cite{micah_allen_0001f13d}\item 

\textbf{Explainable AI (XAI) Systems:} Implement XAI systems to provide clear explanations of algorithmic decisions, fostering transparency and trust in automated processes\cite{leilani_h__gilpin_18bcdd89}.\item 

\textbf{Privacy-Enhancing Technologies (PETs):} Incorporate PETs to minimize personal data exposure and enhance data security, supporting privacy by design and by default principles\cite{may_fen_gan_c61a05ba}.
\end{itemize}

\subsection{Organizational Culture}
At the heart of ethical HDI are practices that prioritize human values and well-being. This involves training, stakeholder engagement, and the creation of channels for feedback and participation.
\begin{itemize}
\item 

\textbf{Training and Development: }Implement comprehensive training programs for staff that cover data ethics, privacy, security, and bias mitigation. This training should be tailored to different roles within the organization and updated regularly to reflect new developments and insights \cite{mubashir_arain_2414e8ff}\item 

 \textbf{User Education and Empowerment: }Develop initiatives to educate users and affected communities about their data rights and how to exercise them. This includes providing accessible resources and support for understanding data practices and the implications of data decisions.\cite{bryce_goodman_a85f1482}
\end{itemize}

\subsection{Stakeholder Engagement:} Create systems for gathering meaningful input and feedback from a diverse range of stakeholders, such as employees, users, and external communities. This involvement should guide the development of policies, the selection of technologies, and the implementation of operational practices, ensuring a variety of perspectives are considered \cite{chrysanthos_dellarocas_21a98b05}. 

Additionally, engage a wide array of stakeholders, including data subjects, employees, regulatory bodies, and civil society organizations, to understand their concerns, expectations, and viewpoints on data usage and ethical matters \cite{brent_mittelstadt_68deb1fc}. This inclusive approach ensures that different viewpoints are integrated into the HDI framework.

\subsection{Data Quality Management}
The structural components of the HDI framework offer a comprehensive blueprint for organizations to navigate the complex ethical challenges inherent in data interaction. By deeply integrating these core elements into their day-to-day operations, organizations can foster a strong culture of responsibility, accountability, and transparency \cite{Swift2001Trust}. 

This, in turn, enhances trust and credibility with a diverse range of stakeholders, from employees to customers to affected communities. Ultimately, the implementation of this ethical HDI framework can enable organizations to make meaningful contributions towards building a more equitable, inclusive, and socially just digital ecosystem that prioritizes human values and well-being.

The implementation of HDR frameworks is a complex undertaking that requires meticulous attention to both ethical and practical considerations. As organisations endeavour to harmonise their data practices with the tenets of HDR, they confront a multitude of challenges and opportunities. The following section delineates pivotal ethical and practical considerations to assist organisations in the responsible integration of HDR frameworks, drawing upon interdisciplinary research to inform these concepts.

\section{Ethical Considerations}
\subsubsection{Respect for Individual Autonomy}
Ensuring that individuals have meaningful control over their data requires robust mechanisms for informed consent and ongoing consent management. Organizations must carefully balance the need for data utility with respect for individual choice and privacy.\cite{maxwell_zostant_786f4300} This involves developing dynamic consent platforms that empower users to adjust their privacy preferences as the contexts and applications of their data evolve over time.

These consent platforms should provide users with clear and accessible information about how their data will be collected, used, and shared. They should also offer granular controls that allow individuals to selectively authorize or revoke access to specific data types or use cases, rather than relying on a one-size-fits-all approach.\cite{Sujata_e7534125}

By giving users the ability to monitor and manage their data preferences on an ongoing basis, organizations can foster a stronger sense of trust and transparency, while still preserving the valuable insights that can be derived from aggregated data. This balanced approach is crucial for upholding individual autonomy while also enabling the responsible and ethical use of data to drive innovation and societal progress.
\subsubsection{Equity and Justice}

Data practices can inadvertently reinforce or exacerbate existing social, economic, and political inequalities. It is crucial for organizations to proactively identify and mitigate various types of biases that may be present in their data collection, analysis, and application processes \cite{betsy_anne_williams_0af5bc10}

 These biases can manifest in the form of underrepresentation of marginalized groups, flawed algorithms that amplify historical discrimination, and the exclusion of diverse perspectives during the development of data-driven systems. To address these challenges, organizations should adopt inclusive design and engagement practices that actively involve a wide range of stakeholders, including underrepresented and marginalized communities, in the development, testing, and ongoing evaluation of their data systems. 

This collaborative approach can help uncover and address blind spots, ensure that the needs and concerns of diverse users are prioritized, and ultimately prevent the perpetuation of harm to vulnerable populations. By embedding equity and justice as core principles throughout the data lifecycle, organizations can work towards building more inclusive, fair, and responsible data practices that benefit all members of society.
\subsubsection{Transparency and Accountability}

Achieving transparency in complex data systems is undoubtedly challenging, but it is essential for ensuring meaningful accountability. Stakeholders, including employees, users, and the broader public, should have access to comprehensive, understandable information about how their data is collected, used, and shared by organizations\cite{federal_data_strategy_df6b7adf}. To address this critical need, organizations must develop robust, user-friendly data policies and communication channels that clearly and transparently inform individuals about the full scope of their data practices. 

This includes providing detailed explanations of the specific purposes for which data is collected and used, as well as the concrete measures and safeguards in place to ensure responsible data management and ethical stewardship of sensitive information\cite{julia_stoyanovich_229ad2db}.

Embracing a high degree of transparency allows organizations to build stronger trust relationships with their stakeholders, empowering individuals to make well-informed choices regarding the use of their personal data. This demonstrates the organization's commitment to upholding data privacy principles and ethical data governance practices.
\subsubsection{Responsible Innovation}

While data-driven technologies offer significant potential for societal benefit, their development and deployment must be guided by strong ethical principles\cite{paul_mccullagh_451e5852}. Establishing comprehensive ethical review processes and frameworks is crucial for enabling responsible innovation. These frameworks should thoroughly assess the far-reaching implications of new data-driven innovations, carefully considering their impact on human and societal well-being \cite{floridiluciano_f31ccd01}

Furthermore, developing and deploying explainable AI tools and practices is essential to ensure transparency and accountability in the application of these transformative technologies. 

Enhancing the accessibility and comprehensibility of algorithmic decision-making processes for nonexpert stakeholders can enable organizations to foster increased trust and facilitate meaningful engagement with these systems. This level of openness and understanding is vital for upholding the principles of ethical data governance and ensuring that the benefits of data-driven innovation are equitably distributed throughout society \cite{olivia_varley_winter_88d5d97f}

The responsible development and application of data-driven technologies must be based on a strong commitment to the protection of ethical principles, the protection of individual rights and the promotion of greater social well-being\cite{luciano_floridi_c93c6698}. Only through this holistic, principle-driven approach can organizations harness the immense potential of data-driven innovation while mitigating the risks and unintended consequences that may arise.
\subsubsection{Environmental and Social Impact }

Responsible data management requires organizations to adopt a holistic, sustainability-focused approach that considers the environmental and social impacts of their data production, storage, and processing activities. This includes carefully examining the energy consumption, emissions, and e-waste generated by their data infrastructure, as well as the broader societal implications of their data-driven systems \cite{carole_jean_wu_b9d1e528}

Adopting a comprehensive perspective on the environmental and societal implications of their data-related activities enables organizations to formulate strategies that mitigate these impacts and foster more sustainable and ethically-grounded data management practices\cite{federica_lucivero_f37ee34f}. This holistic approach is essential to ensure that the advantages of data-driven innovation are balanced against the imperative to protect the planet and address the concerns of diverse stakeholder groups.

\section{Practical Considerations}
\subsection{Integration with Existing Systems}

Integrating HDI principles into an organization's existing data infrastructure and practices requires careful and comprehensive planning to avoid potential disruptions and ensure seamless compatibility\cite{prakash_m_nadkarni_a0ed8639}. Conducting a thorough systems analysis is crucial to identify and address all potential integration challenges. 

This analysis should examine the organization's current data management workflows, technological capabilities, and existing policies and procedures. Based on the insights gained, a well-crafted, phased implementation plan can be developed, allowing for a gradual and strategic adaptation of HDI principles. This measured approach, with clearly defined milestones and feedback loops, enables the organization to mitigate risks, build organizational buy-in, and ensure a smooth transition that preserves data utility and operational continuity. 

Carefully examining the existing data environment and devising a measured, phased implementation plan enables the organization to successfully integrate HDI principles into its fundamental data practices without sacrificing efficiency or effectiveness.
\subsubsection{Technological Feasibility}

Implementing HDI frameworks may necessitate the development or adoption of novel technologies that can effectively address the unique requirements and complexities of data management and interaction in the modern digital landscape. This may include the integration of privacy-preserving data management tools, which employ advanced cryptographic techniques and data anonymization methods to safeguard individual privacy while preserving the utility of data for authorized purposes\cite{security_with_privacy_fdbd6b16}. Furthermore, the implementation of dynamic consent platforms can empower users with greater control and agency over the use of their personal data, allowing them to adjust their preferences as the contexts and applications of data evolve.

Evaluating the feasibility and scalability of these technological solutions is essential to ensure their successful integration within the organization's data infrastructure and practices. This assessment should consider factors such as technical capabilities, ease of implementation, compatibility with existing systems, and the ability to adapt to changing organizational and regulatory requirements over time.\cite{peiran_gao_871e7860} 
\subsubsection{Scalability and Flexibility}

Organizations must design HDI criteria that are highly scalable and adaptable, ensuring their long-term resilience and responsiveness to evolving technologies, regulatory environments, and organizational needs. This entails developing modular frameworks with interchangeable components that can be easily updated or replaced as required, without disrupting the overall system.\cite{ahm_shamsuzzoha_ca962265}

By embracing a modular architecture, organizations can imbue their HDI frameworks with the capacity to adapt rapidly to evolving circumstances. This allows the organization, for instance, to update the pertinent data privacy management components in response to the introduction of new regulatory requirements, without the need to restructure the entire HDI system.

Similarly, as new data analytics tools or interaction techniques emerge, the modular design allows for the seamless integration of these innovations, enabling the organization to stay at the forefront of technological advancements.\cite{sendil_ethiraj_43a013e9}

This scalable and adaptable architecture is crucial for safeguarding the organization's long-term viability and ensuring that its HDI practices remain aligned with evolving needs and best practices. It empowers the organization to proactively anticipate and respond to change, rather than being beholden to rigid, inflexible systems that quickly become obsolete. By prioritizing modularity and adaptability in the design of their HDI frameworks, organizations can future-proof their data practices and maintain a competitive edge in an ever-changing digital landscape.
\subsubsection{Education and Training}

Successful integration of HDI principles necessitates a well-versed and empowered workforce capable of navigating the intricacies of contemporary data systems while upholding ethical data practices. Accordingly, organizations should develop comprehensive training programs for all employees, emphasizing the ethical considerations underlying their work and providing ongoing education on emerging data-related practices and technologies\cite{sean_valentine_dc90b567}. These training programs should not only impart technical knowledge but also foster a deep understanding of the ethical implications of data use, data privacy, and algorithmic decision-making.

Equipping employees with the necessary skillsets and ethical knowledge empowers them to make well-informed, values-aligned decisions when interacting with complex data systems. Additionally, these training initiatives should be accompanied by the establishment of clear organizational policies, accountability frameworks, and channels for continuous feedback and improvement \cite{gopesh_anand_0dafd102}

\section{Implementation Roadmap}
The implementation of a comprehensive Human-Data Interaction (HDI) framework within an organization represents a paradigm shift in data management practices. This roadmap delineates a structured, yet flexible approach to embedding HDI principles into the core of organizational operations. The process is inherently iterative and adaptive, recognizing the dynamic nature of both technological advances and ethical considerations in the digital age.

\subsection{Phase 1: Assessment and Planning}

\subsubsection{Comprehensive Audit and Gap Analysis}

The initial phase commences with a thorough, multidimensional audit of existing data practices. This audit goes beyond simple compliance checks, delving into the ethical implications of current data handling processes. Organizations must scrutinize their entire data ecosystem, from collection points to disposal mechanisms, evaluating each stage against HDI principles \cite{pascale_lehoux_e32b58b4}.

Key components of this audit include:

\begin{itemize}
    \item \textbf{Data flow mapping:} Tracing the journey of data through organizational systems, identifying potential vulnerabilities and ethical pressure points.
    \item \textbf{Policy evaluation:} Assessing the alignment of existing policies with HDI principles, paying particular attention to areas such as consent mechanisms, data minimization, and purpose limitation.
    \item \textbf{Technological infrastructure assessment:} Evaluating current technological capabilities against the requirements of HDI implementation, including privacy-enhancing technologies and explainable AI systems.
    \item \textbf{Stakeholder impact analysis:} Examining how current data practices affect various stakeholders, including employees, customers, and broader communities.
\end{itemize}

This comprehensive audit serves as the foundation for a detailed gap analysis, highlighting areas where current practices fall short of HDI standards and identifying opportunities for improvement.

\subsubsection{Stakeholder Engagement and Participatory Design}

Concurrent with the audit process, organizations must initiate a robust stakeholder engagement program. This goes beyond traditional consultation models, embracing a participatory design approach that actively involves stakeholders in shaping HDI implementation strategies.

Key stakeholder engagement activities include:

\begin{itemize}
    \item \textbf{Multi-stakeholder workshops:} Facilitating interactive sessions that bring together diverse perspectives, including data subjects, employees, regulatory bodies, and civil society organizations.
    \item \textbf{Ethical deliberation forums:} Creating spaces for in-depth discussions on the ethical implications of data practices, encouraging stakeholders to grapple with complex trade-offs and scenarios.
    \item \textbf{User experience research:} Conducting detailed studies to understand how different stakeholder groups interact with and perceive data systems, informing the design of more human-centric interfaces and processes.
\end{itemize}

This participatory approach not only enriches the implementation strategy, but also fosters a sense of shared ownership and commitment to HDI principles across the stakeholder ecosystem.

\subsubsection{Strategic Alignment and Roadmap Development}

Building on the insights from the audit and stakeholder engagement, organizations must align their strategic objectives with HDI principles. This process involves:

\begin{itemize}
    \item \textbf{Vision articulation:} Crafting a clear, compelling vision for how HDI will transform the organization's approach to data management and utilization.
    \item \textbf{Goal setting:} Establishing specific, measurable, achievable, relevant, and time-bound (SMART) goals for HDI implementation.
    \item \textbf{Resource allocation:} Identifying and securing the necessary resources, including budget, personnel, and technological infrastructure, to support HDI implementation.
    \item \textbf{Phased implementation planning:} Developing a detailed, phased plan for rolling out HDI initiatives, prioritizing high-impact areas while managing organizational change effectively.
\end{itemize}

This strategic alignment process ensures that HDI implementation is not treated as a peripheral initiative but is integrated into the core of organizational strategy and operations \cite{sandra_waddock_52778a31}.

\subsection{Phase 2: Policy Development and Technological Integration}
The second phase of HDI implementation marks a critical transition from conceptual understanding to practical application. This phase focuses on translating the HDI principles into tangible organizational policies and technological solutions. It represents a pivotal moment where abstract ethical considerations are transformed into concrete operational practices.

Policy development in this context goes beyond mere regulatory compliance, with the aim of integrating HDI principles into the fabric of organizational decision making. Simultaneously, technological integration involves not just the adoption of new tools, but a fundamental reimagining of data systems to support human-centric interactions.

This phase demands a delicate balance between ethical imperatives and operational realities, requiring unprecedented collaboration across diverse organizational functions. It requires a flexible approach that can adapt to the rapidly evolving digital landscape while maintaining a steadfast commitment to HDI principles \cite{richard_mortier_ab4c0698,floridi2018ai4people}.

Success in this phase sets the foundation for a transformative approach to data management, positioning organizations at the forefront of ethical, human-centered data practices in the digital age.

\subsubsection{Comprehensive Policy Reformation}

The second phase focuses on translating HDI principles into concrete policies and procedures. This process goes beyond simply updating existing policies, often requiring a fundamental reimagining of the organization's approach to data governance.

Key aspects of policy development include:

\begin{itemize}
    \item \textbf{Ethical framework establishment:} Creating a clear, actionable ethical framework that guides all data-related decision-making within the organization.
    \item \textbf{Data lifecycle policies:} Developing comprehensive policies that address each stage of the data lifecycle, from collection to disposal, ensuring HDI principles are embedded throughout.
    \item \textbf{Consent and control mechanisms:} Designing policies that empower individuals with meaningful control over their data, including granular consent options and easy-to-use data management interfaces.
    \item \textbf{Accountability structures:} Establishing clear lines of responsibility and accountability for HDI implementation, including the creation of new roles or committees dedicated to ethical data oversight.
\end{itemize}

These policies should be developed through a collaborative process, involving legal experts, ethicists, technologists, and representatives of various organizational departments to ensure comprehensive coverage and practical applicability \cite{b__tyr_fothergill_8106b012}.

\subsubsection{Technological Infrastructure Enhancement}

Concurrent with policy development, organizations must undertake a significant enhancement of their technological infrastructure to support HDI principles. This involves:

\begin{itemize}
    \item \textbf{Privacy-Enhancing Technologies (PETs):} Implementing advanced PETs, such as differential privacy, secure multi-party computation, and homomorphic encryption, to protect individual privacy while maintaining data utility.
    \item \textbf{Consent Management Platforms:} Developing or adopting sophisticated consent management systems that allow for dynamic, context-aware consent processes.
    \item \textbf{Explainable AI Systems:} Integrating explainable AI technologies to enhance the transparency and interpretability of algorithmic decision-making processes.
    \item \textbf{Data Provenance and Auditing Tools:} Implementing robust systems for tracking data lineage and enabling comprehensive audits of data usage and transformations.
\end{itemize}

The selection and implementation of these technologies should be guided by a thorough assessment of their alignment with HDI principles and their potential impact on stakeholder experiences.

\subsubsection{Organizational Capacity Building}

To support the successful implementation of new policies and technologies, organizations must invest heavily in building internal capacity for HDI. This includes:

\begin{itemize}
    \item \textbf{Comprehensive Training Programs:} Developing and delivering tailored training programs for different organizational roles, covering both technical aspects of HDI implementation and ethical decision-making in data-driven contexts.
    \item \textbf{Change Management Initiatives:} Implementing structured change management processes to facilitate the cultural and operational shifts required for HDI adoption.
    \item \textbf{Skills Development:} Investing in upskilling and reskilling programs to ensure staff have the necessary competencies to operate effectively within an HDI framework.
    \item \textbf{Cross-functional Collaboration:} Fostering environments and processes that encourage collaboration between different organizational functions, breaking down silos that may impede HDI implementation.
\end{itemize}

These capacity-building efforts are crucial for ensuring that HDI principles are not just understood theoretically but are actively applied in day-to-day operations across the organization \cite{upol_ehsan_edcb5aa7}.

\subsection{Phase 3: Implementation and Evaluation}

\subsubsection{Phased Rollout and Iterative Refinement}

The implementation phase should follow a carefully structured, phased approach:

\begin{itemize}
    \item \textbf{Pilot Programs:} Initiating small-scale pilot programs in selected departments or processes to test HDI implementations in controlled environments.
    \item \textbf{Feedback Loops:} Establishing robust mechanisms for gathering and analyzing feedback from stakeholders involved in pilot programs.
    \item \textbf{Iterative Refinement:} Using insights from pilot programs to refine policies, technologies, and processes before broader rollout.
    \item \textbf{Scaled Implementation:} Gradually expanding HDI implementations across the organization, maintaining flexibility to adapt to emerging challenges and opportunities.
\end{itemize}

This phased approach allows organizations to learn from their first experiences and make the necessary adjustments, increasing the likelihood of a successful adoption throughout the organization.

\subsubsection{Continuous Monitoring and Adaptive Governance}

As the principles of HDI become more widely adopted, organizations need to set up thorough monitoring and governance frameworks:

\begin{itemize}
    \item \textbf{Real-time Monitoring Systems:} Implementing technologies and processes for continuous monitoring of HDI compliance and effectiveness across all data operations.
    \item \textbf{Stakeholder Feedback Mechanisms:} Creating multiple channels for ongoing stakeholder input, including regular surveys, focus groups, and open feedback platforms.
    \item \textbf{Ethical Review Boards:} Establishing cross-functional ethical review boards to assess complex cases and guide policy evolution.
    \item \textbf{Adaptive Governance Frameworks:} Developing flexible governance structures that can evolve in response to new ethical challenges, technological advancements, and regulatory changes.
\end{itemize}

These monitoring and governance mechanisms ensure that HDI implementation remains responsive to changing contexts and emerging ethical considerations.

\subsubsection{Impact Assessment and Reporting}

Regular, comprehensive impact assessments are crucial to understanding the effectiveness of HDI implementations and identifying areas for improvement.

\begin{itemize}
    \item \textbf{Multidimensional Impact Metrics:} Developing a set of metrics that capture the multifaceted impacts of HDI implementation, including ethical, social, economic, and operational dimensions.
    \item \textbf{Longitudinal Studies:} Conducting long-term studies to track the evolving impacts of HDI practices on various stakeholder groups and organizational outcomes.
    \item \textbf{Transparent Reporting:} Producing regular, detailed reports on HDI implementation progress, challenges, and impacts, making these accessible to all stakeholders.
    \item \textbf{External Audits:} Engaging independent third parties to conduct periodic audits of HDI practices, ensuring objectivity and credibility in evaluations.
\end{itemize}

These assessment and reporting practices not only drive continuous improvement but also contribute to building trust and accountability with stakeholders.

\subsection{Phase 4: Adaptation and Evolution}

\subsubsection{Horizon Scanning and Proactive Adaptation}

To remain effective in a rapidly evolving digital landscape, organizations must develop robust capabilities to anticipate and respond to emerging trends and challenges.

\begin{itemize}
    \item \textbf{Technology Foresight:} Establishing dedicated teams or processes for monitoring and analyzing emerging technologies and their potential ethical implications.
    \item \textbf{Regulatory Tracking:} Developing systems for staying abreast of evolving data protection regulations and proactively adapting HDI practices to ensure compliance.
    \item \textbf{Ethical Foresight:} Engaging in ongoing ethical deliberation to anticipate future ethical challenges arising from new data uses or societal changes.
    \item \textbf{Collaborative Innovation:} Participating in industry collaborations, academic partnerships, and multi-stakeholder initiatives to collectively address emerging HDI challenges.
\end{itemize}

This proactive stance enables organizations to stay ahead of the curve, adapting their HDI practices to address new challenges before they become critical issues.

\subsubsection{Continuous Learning and Knowledge Management}

Building a robust knowledge management system around HDI practices is essential for long-term success:

\begin{itemize}
    \item \textbf{Case Study Development:} Documenting and analyzing specific cases of HDI implementation, including successes, challenges, and lessons learned.
    \item \textbf{Best Practice Repositories:} Creating and maintaining accessible repositories of HDI best practices, updated regularly with new insights and experiences.
    \item \textbf{Cross-organizational Learning:} Facilitating knowledge sharing across different parts of the organization to ensure consistent application of HDI principles.
    \item \textbf{External Knowledge Exchange:} Engaging in knowledge exchange with other organizations, industry bodies, and academic institutions to broaden the collective understanding of effective HDI practices.
\end{itemize}

These learning and knowledge management practices ensure that the organization continuously builds on its HDI capabilities, fostering a culture of ongoing improvement and innovation.

\subsubsection{Long-term Strategic Integration}

Finally, organizations must work towards fully integrating HDI principles into their long-term strategic planning:

\begin{itemize}
    \item \textbf{Strategic Visioning:} Regularly revisiting and refining the organization's long-term vision for ethical data management, ensuring it remains aligned with evolving HDI principles and societal expectations.
    \item \textbf{Innovation Alignment:} Ensuring that all data-driven innovation initiatives are evaluated through an HDI lens from their inception, making ethical considerations a fundamental part of the innovation process.
    \item \textbf{Organizational Culture Evolution:} Continuously working to embed HDI principles into the core values and culture of the organization, making ethical data practices an intrinsic part of organizational identity.
    \item \textbf{Stakeholder Relationship Evolution:} Developing more collaborative, trust-based relationships with stakeholders, positioning the organization as a leader in ethical data practices.
\end{itemize}

This long-term strategic integration ensures that HDI is not treated as a separate initiative, but becomes an integral part of how the organization operates and creates value in the digital age.

By following this comprehensive and in-depth roadmap, organizations can embark on a transformative journey towards ethical, human-centered data practices. This approach not only ensures compliance with evolving regulations but positions organizations at the forefront of responsible innovation, building trust and creating sustainable value in an increasingly data-driven world.

\section{Enhancing HDI Culture}
For an organization to truly embody HDI principles, a deep-rooted understanding of these concepts among all members is vital. Regular training sessions should be conducted to educate employees about the ethics of data use, responsible data handling, and the potential risks associated with improper data management. Furthermore, it's important to extend educational efforts to stakeholders, including users and affected communities, thereby increasing their data literacy. This empowers them to participate actively and effectively in data governance processes, ensuring their voices are heard and considered in decision-making.

However, cultivating an HDI-centric culture within an organization is not a one-time achievement but a continuous process. It requires the establishment of mechanisms for ongoing monitoring, feedback, and iterative improvements. Such processes are crucial for maintaining the effectiveness of HDI practices amidst the rapidly changing dynamics of data use and technology \cite{anastasija_nikiforova_d22daaaa}

By committing to constant monitoring and adaptation, an organization showcases its dedication to continuously enhancing its HDI practices. This proactive approach not only fosters trust among stakeholders but also positions the organization as a frontrunner in responsible and ethical data management. As the digital environment evolves, staying ahead in implementing robust HDI frameworks becomes a testament to an organization's commitment to ethical data use and governance.

Training and education form the cornerstone of implementing HDI frameworks effectively. This comprehensive approach is key to building a culture that values ethical data use, adheres to HDI principles, and equips both employees and users with the necessary skills and knowledge to navigate the complexities of contemporary data ecosystems. The outlined strategies underscore the importance of developing targeted training and education programs that are integral to fostering an environment where HDI principles are not just understood but actively practiced.

\subsection{Employee Training Programs}
Equip employees with a deep understanding of HDI principles, including data ethics, responsible data handling, and awareness of biases. To further empower individuals as active stakeholders in the data ecosystem, the HDI Framework calls for proactive user education initiatives \cite{sanjay_kumar_singh_01559ecb}. This training should extend beyond theoretical knowledge, emphasizing practical skills for implementing HDI in their daily responsibilities.

Strategies:
\begin{itemize}
\item 

\textbf{Curriculum Development:} Design a curriculum that covers essential topics such as data privacy, consent protocols, algorithmic transparency, and bias mitigation. This curriculum should be adaptable, allowing for updates as new ethical challenges and technological developments arise \cite{tiago_palma_pagano_16b9fa15}\item 

\textbf{Interactive Learning:} Employ interactive learning techniques, such as case studies, simulations, and role-playing exercises. These methods encourage active engagement and help employees internalize HDI concepts by applying them to real-world scenarios\cite{authentic_learning_environments_in_higher_education_8d6a51ce,jim_x__chen_24b1d11c}.\item 

\textbf{Cross-Functional Training: }Implement cross-functional training sessions to foster interdisciplinary understanding and collaboration\cite{ruth_stock_da2c6801}. Exposing employees to interdisciplinary perspectives and diverse challenges across different organizational departments can foster a comprehensive approach to implementing HDI principles.
\end{itemize}

\subsection{Stakeholder Education}
Empower users and affected communities by enhancing their understanding of data rights and the implications of data practices. Education programs should aim to demystify data operations and foster informed participation in data ecosystems.

Strategies:
\begin{itemize}
\item 

\textbf{Accessible Materials: }Develop educational materials that are accessible to non-experts, including infographics, videos, and simple guides. These resources should explain users' data rights, how their data is used, and ways they can control their personal information\cite{paul_de_hert_73ead814}.\item 

\textbf{Community Engagement:} Organize workshops and forums in collaboration with community organizations. These events can serve as platforms for dialogue, where users can express concerns, ask questions, and provide feedback on data practices\item 

\textbf{Feedback Mechanisms:} Establish mechanisms for users to provide ongoing feedback on data practices and education materials. This feedback loop is essential for ensuring that education programs remain relevant and responsive to users' needs.
\end{itemize}

\subsection{Continuous Learning and Adaptation}
 Foster an environment of continuous learning and adaptation to address the dynamic nature of data ethics and technology. Training and education programs should evolve in response to new challenges and insights.

Strategies:
\begin{itemize}
\item 

\textbf{Regular Program Updates: }Update training and education programs regularly to incorporate the latest research findings, ethical debates, and technological advancements. This ensures that all stakeholders remain informed about current best practices.\cite{misti_ault_anderson_a82210b6}\item 

 \textbf{Professional Development Opportunities:} Offer professional development opportunities for employees to deepen their expertise in data ethics and HDI-related fields. This could include sponsoring attendance at conferences, online courses, and certification programs.\cite{suzanna_conrad_d9815184}\item 

\textbf{Evaluation and Feedback:} Implement robust evaluation processes to assess the effectiveness of training and education programs. Use feedback from participants to make iterative improvements, ensuring that learning objectives are met.\cite{katie_shilton_287b89c9,diane_o__du_et_46106f6e}
\end{itemize}

\subsection{Cross-functional Collaboration:}
Implementing HDI principles effectively requires a collaborative effort that spans various departments and expertise within an organization. Cross-functional collaboration is essential for addressing the multifaceted challenges of ethical data management and for ensuring that HDI practices are consistently applied across all organizational processes.
Strategies for fostering cross-functional collaboration in HDI implementation include:

\begin{itemize}
    \item Interdisciplinary Teams: Form teams that bring together professionals from diverse backgrounds such as data science, legal, ethics, user experience, and business operations. This diversity ensures that HDI considerations are viewed from multiple perspectives, leading to more comprehensive and nuanced solutions \cite{trist1983referent}.
    \item Collaborative Workshops: Organize regular workshops where team members from different departments can share insights, discuss challenges, and collaboratively develop solutions to HDI-related issues. These sessions can help break down silos and foster a shared understanding of HDI principles \cite{edmondson2018cross}.
    \item Shared Responsibility: Establish HDI as a shared responsibility across all departments, rather than confining it to a single team or unit. This approach ensures that HDI principles are integrated into all aspects of the organization's operations \cite{de2009exploratory}.
    \item Cross-departmental Training: Develop training programs that bring together employees from different departments to learn about HDI principles and practices. This not only enhances knowledge sharing but also builds relationships across functional boundaries \cite{bauer2007workplace}.
    \item Collaborative Tools and Platforms: Implement collaborative tools and platforms that facilitate communication and knowledge sharing about HDI practices across different teams and departments \cite{anders2016team}.
    \item Rotational Programs: Consider implementing rotational programs where employees can spend time in different departments to gain a holistic understanding of how HDI principles apply across various organizational functions \cite{campion1994career}.
\end{itemize}

By fostering cross-functional collaboration, organizations can ensure a more holistic and effective implementation of HDI principles. This collaborative approach not only enhances the quality of HDI practices but also promotes a culture of shared responsibility for ethical data management across the entire organization.

The development and implementation of training and education programs are foundational to the success of HDI frameworks. By prioritizing comprehensive, accessible, and continuously updated training for both employees and users, organizations can cultivate a culture of responsible data stewardship. This not only aligns with ethical imperatives but also enhances trust and cooperation across the data ecosystem, paving the way for a future where data practices are transparent, fair, and human-centered.

\section{Data Lifecycle Management}
Data Lifecycle Management (DLM) is a crucial component of the HDI Framework, ensuring that ethical considerations and human-centric practices are applied at every stage of data handling. This comprehensive approach addresses the entire journey of data from its creation to its eventual disposal, aligning each phase with HDI principles.

\subsection{Collection}
The data collection phase must prioritize ethical practices and user empowerment. Organizations should implement transparent mechanisms for obtaining informed consent, clearly communicating the purpose and scope of data collection. Privacy-by-design principles should be embedded in the collection process, ensuring that only necessary data is gathered \cite{Edwards2018Enslaving}.

\subsection{Processing}
Data processing should adhere to the principles of fairness, transparency, and accountability. Implementing explainable AI techniques can help make algorithmic decision-making processes more transparent and understandable to users \cite{gunning2019xai}. Regular audits of data processing systems should be conducted to identify and mitigate potential biases.

\subsection{Storage}
Secure and responsible data storage is paramount. Organizations must implement robust security measures to protect data from unauthorized access or breaches. Data minimization principles should be applied, storing only essential information for the required duration. Clear policies on data retention and deletion should be established and communicated to users \cite{Teague2018Retention}.

\subsection{Utilization}
The utilization of data should be guided by ethical considerations and respect for individual rights. Organizations should implement mechanisms that allow users to understand how their data is being used and provide options for controlling its use. Ethical review processes must be in place for new data utilization projects, especially those involving sensitive information or potentially high impact decisions \cite{Ferretti2021Ethics}.

\subsection{Sharing}
Data sharing practices must prioritize user consent and control. Clear policies should govern the circumstances under which data can be shared, with whom, and for what purposes. Implement granular consent mechanisms that allow users to selectively share their data and revoke access as needed \cite{Jaiman2020A}.

\subsection{Disposal}
The final stage of the data lifecycle, disposal, must be handled with care. Organizations should establish clear protocols for securely deleting data when it is no longer needed or when users request its removal. This process should be verifiable and in compliance with relevant data protection regulations \cite{Reardon2014On}.

By meticulously managing each stage of the data lifecycle in alignment with HDI principles, organizations can foster a more ethical, transparent, and user-centric approach to data handling. This comprehensive lifecycle management not only enhances compliance with regulatory requirements but also builds trust with users and stakeholders, ultimately contributing to a more responsible and sustainable data ecosystem.

\section{Monitoring and Adaptation}
The implementation of HDI principles within an organization is an ongoing, dynamic process that requires constant vigilance and flexibility to adapt to the ever-changing landscape of regulations, technological advancements, and societal expectations. To maintain and enhance the effectiveness of HDI practices, organizations must establish robust mechanisms for continuous monitoring, regular reassessment, and agile adaptation of practices \cite{johannes_m__pennings_6de3e29e}.

\subsection{Continuous Assessment}

Organizations should implement structured processes for regularly reviewing their data handling and interaction policies. This involves not only examining the internal mechanisms of data management and governance but also actively seeking feedback from a wide range of stakeholders, including employees, users, and external communities affected by the organization's data practices \cite{lisa_r__hirschhorn_88a0bf10}. Techniques such as employee surveys, user satisfaction assessments, and the establishment of a dedicated data ethics committee can provide invaluable insights into the efficacy of existing practices and highlight areas requiring improvement or adjustment.

\subsection{Feedback Mechanisms}

Establishing comprehensive feedback channels that enable stakeholders to actively report concerns, provide suggestions, and share their perspectives is crucial. This continuous feedback loop is essential for driving iterative improvements to the organization's data governance policies, technological solutions, and overall HDI implementation \cite{christian_moltu_53e8bb34}.

\subsection{Adaptive Governance}

As the data landscape evolves, so too must the policies and practices that govern it. Organizations should develop flexible governance structures that can evolve with changing needs, regulatory requirements, and technological capabilities. This may involve regular policy reviews, updates to data management protocols, and adjustments to decision-making processes \cite{andrew_georgiou_5a866d90}.

\subsection{Emerging Technology Integration}

Stay informed about emerging trends, technologies, and regulatory developments that could impact ethical data practices. Continuously research and analyze the implications of new advancements in areas such as privacy-enhancing technologies, consent management platforms, and explainable AI systems. Adapt the HDI framework to proactively incorporate these advancements, ensuring that the organization remains at the forefront of ethical data interaction \cite{p__jonathon_phillips_a628cb50}.

\subsection{Performance Metrics}

Develop quantifiable metrics to gauge the performance of HDI initiatives. These might include user consent rates, the volume of data access or modification requests, incident reports relating to data security breaches, and evaluations of algorithmic bias. Incorporating these metrics into visual dashboards can facilitate broader organizational awareness and engagement with HDI performance indicators, bridging the gap between technical and non-technical stakeholders \cite{pedro_elias_lucas_ramos_meireles_4b7ef0e4}.

\subsection{External Audits}

Carry out regular independent evaluations of HDI practices. Third-party audits can offer an unbiased review of the organization's compliance with HDI principles, unveil possible blind spots, and suggest new perspectives for enhancement \cite{andrew_georgiou_5a866d90}. By cultivating a culture of continuous improvement and adaptability, organizations can maintain and develop their HDI practices over time. This strategy guarantees that HDI practices stay relevant and responsive to the requirements and issues of all stakeholders, thereby preserving trust and confidence in the organization's data practices and contributing to a more ethical and responsible data-driven society \cite{johannes_m__pennings_6de3e29e}. 

The successful implementation of HDI relies on an organization's capacity to remain agile, responsive, and dedicated to ethical data practices amid rapid technological and societal changes. By adopting this dynamic approach to monitoring and adaptation, organizations can not only achieve current standards for responsible data management but also predict and tackle future challenges in the constantly evolving digital environment. 

The process of monitoring and adjusting HDI practices is fundamentally linked to the intricate and changing nature of the data ecosystem. As organizations strive to uphold effective HDI frameworks, they must also navigate a landscape marked by complex data systems, emerging technologies, and evolving societal expectations. This balance between ongoing monitoring and the need to address complex, dynamic factors highlights the importance of a comprehensive, forward-looking approach to HDI implementation. 

The following section examines these complex and dynamic factors, discussing how organizations can effectively manage the complexities of data systems, foster an HDI-centric organizational culture, and stay responsive to developing trends and context-specific challenges. By dealing with these factors alongside robust monitoring and adaptation practices, organizations can ensure that their HDI frameworks stay both effective and relevant in the face of a continuously changing digital environment \cite{david_j__hand_687e598b}.

\subsection{Complex and dynamic considerations}

Implementing HDI frameworks within organizations involves navigating a complex landscape of ethical and practical considerations. These considerations are critical for developing data practices that not only comply with technical standards but also align with ethical imperatives and societal expectations.

\subsubsection{Complexity}
Data systems within contemporary organizations exhibit remarkable complexity\cite{mike_hinchey_a10def06}. These systems frequently evolve incrementally, encompass diverse stakeholder interests, and operate across varied technological platforms. Implementing HDI principles in such intricate environments requires a nuanced and strategic approach.

A modular strategy, which focuses on applying HDI principles to specific and well-defined aspects of the data lifecycle, can be effective. Successes in these targeted areas can build organizational confidence and momentum, facilitating a gradual expansion of the HDI principles in broader operational domains. This phased strategy helps manage the risk of becoming overwhelmed by the wide range of possible modifications required for complete HDI integration\cite{jeffrey_d__sachs_32d53b09}.

\subsubsection{Envolving Technologies}
The data landscape is constantly changing, shaped by rapid technological advancements, evolving regulatory environments, and changing societal expectations\cite{harnessing_the_data_revolution_dbdfe8e8}. Organizations must remain vigilant and agile in adapting their HDI frameworks to these dynamic circumstances. 

This necessitates a continuous process of monitoring, evaluation, and refinement, whereby organizations regularly assess the efficacy of their HDI practices and make adjustments as needed.Mechanisms for gathering user feedback, tracking emerging best practices, and proactively scanning for regulatory changes must be embedded within the organization's operational structures. 

By maintaining a pulse on the evolving HDI landscape, organizations can ensure that their data practices remain relevant, effective, and aligned with the best interests of their stakeholders. 

\subsubsection{Regulatory Landscape:}
The regulatory environment surrounding data protection and privacy is in constant flux, with new laws and regulations being introduced or updated regularly. Organizations must stay informed of these changes and adapt their HDI practices accordingly. 

This requires a proactive approach to compliance, involving regular legal reviews and the ability to quickly implement necessary changes to data handling processes. The challenge lies not only in meeting current regulatory requirements but also in anticipating future legislative developments \cite{tikkinen2018eu}.

\subsubsection{Societal Landscape:}
Public attitudes towards data privacy and ethical data use are evolving rapidly, influenced by high-profile data breaches, privacy scandals, and increased awareness of the value and vulnerability of personal data. Organizations must be attuned to these shifting societal expectations and be prepared to adjust their data practices to maintain public trust. This may involve going beyond simple regulatory compliance to embrace more stringent ethical standards that align with public sentiment \cite{auxier2019americans}.

\subsubsection{Ethical Dilemmas:}
As data-driven technologies become more sophisticated, organizations increasingly face complex ethical dilemmas that may not have clear-cut solutions. These could range from questions about the appropriate use of AI in decision-making processes to concerns about the potential misuse of predictive analytics. Addressing these dilemmas requires a robust ethical framework and decision-making process that can navigate the nuanced terrain of data ethics \cite{mittelstadt2016ethics}.

\subsubsection{Cross-border Data Flows:}
In an increasingly globalized digital economy, organizations often need to transfer data across national borders. This presents unique challenges in terms of complying with varying (and sometimes conflicting) data protection laws in different jurisdictions. Implementing HDI principles in this context requires a nuanced understanding of international data protection regulations and the ability to adapt data handling practices to meet diverse legal requirements \cite{kuner2013transborder}.

\subsubsection{Balancing Innovation and Ethics:}
Organizations face the ongoing challenge of balancing the drive for innovation with the need for ethical data practices. While data-driven innovation can lead to significant advancements and competitive advantages, it must be pursued in a way that respects individual privacy and upholds ethical standards. This balance requires careful consideration of the potential impacts of new data uses and technologies, and the development of governance structures that promote responsible innovation \cite{luciano_floridi_c93c6698}.

\subsubsection{Adaptive Strategies: }
Given the rapidly evolving nature of the data landscape, organizations need to develop adaptive strategies that can respond quickly to new challenges and opportunities. This involves creating flexible governance structures, fostering a culture of continuous learning and improvement, and developing the capacity to pivot data practices in response to emerging ethical concerns or technological advancements. Adaptive strategies should also include regular reviews and updates of HDI frameworks to ensure they remain relevant and effective \cite{janssen2016adaptive}.

The complex and dynamic considerations inherent in implementing HDI frameworks inevitably give rise to a host of challenges that organizations must address. As we have seen, factors such as the intricacy of data systems, the need for adaptive organizational cultures, and the rapidly evolving technological landscape create a multifaceted environment in which HDI principles must be operationalized. 

These complexities naturally lead to specific challenges that require innovative solutions. In the following section, we will explore these challenges in detail and propose practical, research-backed solutions to overcome them. By examining issues such as technical integration, ethical decision-making, user empowerment, and regulatory compliance, we can provide a comprehensive roadmap for organizations seeking to navigate the intricate terrain of HDI implementation. 

This approach not only addresses the complexities discussed earlier but also offers tangible strategies for overcoming obstacles, thereby enabling organizations to more effectively embed HDI principles into their core operations \cite{sandra_waddock_52778a31}.

\section{HDI Framework Challenges and Solutions}
Implementing a HDI framework within an organization encompasses a myriad of challenges, ranging from technical complexities to ethical dilemmas. Addressing these challenges effectively requires a multifaceted approach that incorporates both innovative solutions and a commitment to ethical principles. This section outlines key challenges associated with HDI implementation and proposes viable solutions, grounded in scholarly research and best practices.

\subsection{Technical Complexities}
Challenge: Integrating HDI principles into existing data management infrastructure. Organizations often have well-established data pipelines, analytical tools, and storage systems in place.

Solution: Adopt a modular and iterative approach to HDI integration, focusing on specific, well-defined aspects of the data lifecycle.This allows organizations to build momentum and confidence through early successes, before gradually expanding HDI principles across broader operational domains.

\subsection{Data Complexity and Integration}
\textbf{Challenge}: Modern organizations operate within highly complex data ecosystems that have evolved over time. These systems often feature diverse data sources, varying data formats, and legacy systems that may not easily support new HDI principles.

\textbf{Solution}: Adopting a modular, phased approach to implementing HDI principles can help organizations manage the complexity. Initiating small-scale, pilot projects to integrate HDI principles in specific areas can provide valuable insights and allow for gradual expansion across the organization. Leveraging data integration tools and adopting standards for data interoperability can also facilitate smoother integration of HDI principles.

\subsection{Ethical Decision-Making and Bias Mitigation}
\textbf{Challenge}: Ensuring ethical decision-making and mitigating biases in algorithms and datasets represent significant challenges, as these biases can lead to discriminatory outcomes and undermine trust in data system.

\textbf{Solution}: Implementing robust ethical review processes and adopting transparent decision-making frameworks can enhance accountability. Furthermore, integrating bias detection and correction methodologies, such as algorithmic audits and inclusive dataset design, can actively mitigate biases, promoting fairness and non-discrimination in data practices.

\subsection{User Empowerment and Participation}
\textbf{Challenge}: Truly empowering users within the HDI framework requires moving beyond nominal consent mechanisms to enable meaningful user control over their data.

\textbf{Solution}: Developing and deploying consent management platforms that offer users granular control over their data, alongside clear, accessible information about data use practices, can enhance user empowerment. Additionally, creating channels for user feedback and participation in decision-making processes reinforces user agency and ensures that HDI practices align with user expectations and values.

\subsection{Organizational Culture and Leadership}
\textbf{Challenge}: Cultivating an organizational culture that genuinely prioritizes HDI values, such as transparency, accountability, and ethical data use, requires significant shifts in norms and practices.

\textbf{Solution}: Leadership commitment to HDI principles is crucial. This can be demonstrated through resource allocation, setting clear ethical guidelines, and recognizing employees who champion responsible data stewardship. Fostering a culture of continuous learning and ethical reflection, where data ethics are integrated into everyday practices, can embed HDI values deeply within the organizational fabric.

\subsection{Regulatory Compliance and Evolving Standards}
\textbf{Challenge}: The dynamic nature of data regulation and the emergence of new technological capabilities necessitate ongoing vigilance and adaptability to ensure compliance and ethical integrity.

\textbf{Solution}: Establishing dedicated teams to monitor regulatory developments and technological advancements can help organizations stay ahead of compliance requirements and ethical considerations. Engaging in industry consortia and standard-setting bodies can also provide insights into emerging trends and best practices, facilitating proactive adaptation of HDI practices.

\subsection{Cost Implications}
\textbf{Challenge}: The implementation of an HDI framework can entail significant costs, including investments in new technologies, training programs, and potentially restructuring data governance practices. Organizations may face financial constraints that limit their ability to fully adopt HDI principles.

\textbf{Solution}: Strategic planning and prioritization can help mitigate cost concerns. Organizations should assess the most critical areas for HDI integration based on risk, impact, and strategic value, focusing initial investments where they can achieve the most significant benefits. Leveraging open-source technologies and platforms for data management and ethics training can also reduce costs. Additionally, incremental implementation allows for the spreading of costs over time, making the process more financially manageable.

\subsection{Data Literacy Gaps}
\textbf{Challenge}: A fundamental barrier to HDI implementation is the varying levels of data literacy across an organization, from executive leadership to operational staff. Insufficient understanding of data principles can hinder effective decision-making and limit the capacity to engage with HDI tools and policies meaningfully.

\textbf{Solution}: Comprehensive data literacy programs are essential for bridging these gaps. Such programs should cater to different roles within the organization, providing tailored training that ranges from basic data awareness to advanced analytical skills. Furthermore, creating a culture that values continuous learning and curiosity about data can encourage self-driven improvement in data literacy across the organization.

\subsection{Balancing Innovation with Responsibility}
\textbf{Challenge}: Organizations striving to innovate with data often encounter tensions between pushing the boundaries of what is technologically possible and adhering to ethical data practices. This balance is crucial to maintaining trust and ensuring long-term sustainability.

\textbf{Solution}: Establishing clear ethical guidelines that define responsible innovation practices is vital. These guidelines should encourage creativity and exploration while setting firm boundaries to prevent unethical data use. Regular ethical reviews of new projects and technologies can ensure that innovation aligns with HDI principles. Engaging with external ethics experts and stakeholders can also provide diverse perspectives that enrich the organization's approach to balancing innovation with responsibility.

In conclusion, the HDI framework presents a comprehensive model for a future driven by data, where the needs, concerns, and values of humans are central to the design and deployment of data-driven technologies. Implementing an HDI framework is a complex undertaking that requires addressing technical, ethical, and organizational challenges. To navigate these challenges effectively, organizations should adopt a strategic, phased integration approach, commit to ethical principles, and foster a culture of transparency and accountability. By doing so, they can enhance their data practices and contribute to building a more equitable and trustworthy digital ecosystem.

Addressing the cost implications, data literacy gaps, and the need to balance innovation with responsibility is crucial for the successful integration of HDI principles into organizational practices. By adopting strategic, inclusive, and ethical approaches, organizations can navigate these challenges effectively. This not only enhances their capability to implement HDI frameworks but also strengthens their position as leaders in responsible data management and innovation.

\section{Conclusions and Future Works}

The Human-Data Interaction (HDI) Framework presented in this article marks a fundamental paradigm shift in the conceptualization and implementation of data management and utilization practices within organizations. This framework represents a radical departure from traditional data-centric approaches, which often prioritize technological capabilities and organizational efficiency over human concerns. Instead, HDI places human values, rights, and agency at the forefront of data practices, recognizing that data systems exist within a complex socio-technical ecosystem where ethical considerations are paramount.

By centering human needs and experiences, HDI offers a comprehensive model for navigating the intricate ethical landscape of our increasingly data-driven future. This approach acknowledges the profound impact that data practices have on individual lives, societal structures, and the distribution of power in the digital age. It seeks to rebalance the relationship between individuals and data systems, moving away from viewing people as mere data subjects to recognizing them as active agents with inherent rights and dignity.

The HDI framework addresses the multifaceted challenges arising from the pervasive use of data in modern society, including issues of privacy, consent, algorithmic bias, and the opacity of data-driven decision-making processes. It provides a structured approach for organizations to ethically manage the entire data lifecycle, from collection and processing to analysis and application. This holistic perspective enables organizations to anticipate and mitigate potential ethical pitfalls, fostering trust and accountability in their data practices.

Moreover, the HDI framework serves as a bridge between technological innovation and ethical responsibility. It challenges the notion that these two aspects are inherently at odds, instead proposing that true innovation in the data sphere must be grounded in respect for human values. By doing so, it paves the way for a future where technological advancements and ethical considerations evolve in tandem, mutually reinforcing each other to create data ecosystems that are not only powerful and efficient but also just, transparent, and human-centric.

In essence, the HDI framework represents a critical evolution in our understanding of the role of data in society. It offers a roadmap for organizations to navigate the complex ethical terrain of the digital age, ensuring that as we harness the transformative power of data, we do so in a manner that upholds human dignity, promotes societal well-being, and cultivates a more equitable digital future for all.

\subsection{Key Takeaways and Implications}
The Human-Data Interaction (HDI) Framework presented in this article offers a transformative approach to data management and utilization, with significant implications for organizations, individuals, and society. As we synthesize the core elements of this framework, several key takeaways emerge, each providing actionable insights for the practical implementation of ethical data systems.

These takeaways encapsulate the essence of the HDI framework's contribution to data ethics and provide a roadmap for responsible data management in the digital age. They challenge us to rethink the relationship between humans and data, proposing a paradigm where ethical considerations are foundational to data systems.

The implications of these insights extend beyond data management, influencing organizational culture, technological innovation, and societal values. They set the stage for future research and development in HDI, highlighting areas needing further investigation and pointing towards emerging trends in ethical data practices.
In the subsequent points, we will delve into each of these key insights, analyzing their importance, obstacles, and possibilities for fostering a more fair, open, and human-focused data environment.

\begin{itemize}
    \item \textbf{Ethical Foundations:} At the core of the HDI framework are foundational principles including agency, legibility, equity, accountability, privacy, and stakeholder engagement. These principles serve as the ethical bedrock for all data practices, ensuring that human dignity and rights are upheld in increasingly data-centric environments. The implementation of these principles represents a significant departure from traditional data management approaches, necessitating a fundamental reevaluation of how organizations collect, process, and utilize data. This ethical foundation challenges organizations to move beyond compliance-driven approaches to data management and instead embrace a proactive stance that anticipates and addresses ethical concerns throughout the data lifecycle.
    \item \textbf{Holistic Approach:} HDI necessitates a multidisciplinary approach, integrating insights from technology, law, ethics, and user experience design. This holistic perspective enables organizations to address the multifaceted challenges of ethical data management comprehensively. By breaking down silos between different domains of expertise, HDI fosters innovative solutions that are both technically robust and ethically sound. This integration of diverse perspectives is crucial for developing data practices that can withstand scrutiny from various stakeholders and adapt to evolving societal expectations.
    \item \textbf{Structural Implementation:} The framework provides practical guidance for embedding HDI principles into organizational structures, including data governance frameworks, technological infrastructure, and human-centric practices. This structural approach ensures that HDI is not merely aspirational but operationally integrated. The challenge lies in adapting existing organizational structures and processes to accommodate these new ethical imperatives. This may involve creating new roles, such as ethics officers or data stewards, redesigning data workflows to incorporate ethical checkpoints, and developing new metrics for measuring the ethical performance of data systems.
    \item \textbf{Adaptive Strategies:} Given the rapidly evolving nature of technology and data practices, the HDI framework emphasizes the importance of continuous monitoring, feedback, and adaptation. This dynamic approach ensures that HDI practices remain relevant and effective in the face of emerging challenges and opportunities. Organizations must develop the capability to anticipate and respond to changes in the technological, regulatory, and societal landscape. This might involve establishing dedicated teams for horizon scanning, regularly reviewing and updating ethical guidelines, and fostering a culture of ethical reflection and debate within the organization.
    \item \textbf{Cultural Transformation:} Implementing HDI requires a significant shift in organizational culture, emphasizing transparency, accountability, and ethical data stewardship. This cultural change is crucial for the long-term success and sustainability of HDI practices. It involves not only training and education but also a fundamental realignment of organizational values and priorities. Leaders must model ethical behavior, incentivize ethical decision-making, and create an environment where employees feel empowered to raise ethical concerns. This cultural transformation extends beyond the organization to its relationships with customers, partners, and the broader community, fostering a new paradigm of trust and collaboration in the data ecosystem.
\end{itemize}

 By embracing these key takeaways and implications, organizations can begin to realize the full potential of the HDI framework, creating data practices that are not only innovative and efficient but also ethical, transparent, and deeply respectful of human values and rights.

\subsection{Future Research Directions}

As the field of HDI continues to evolve, several key areas warrant further investigation:

\begin{enumerate}
    \item \textbf{Adaptive Ethical Frameworks}:
    \begin{itemize}
        \item Develop dynamic ethical guidelines that can evolve with technological advancements.
        \item Investigate methods for real-time ethical assessment of data practices.
        \item Explore the use of AI in monitoring and enforcing ethical data standards.
    \end{itemize}
    \item \textbf{Advanced Privacy-Enhancing Technologies (PETs)}:
    \begin{itemize}
        \item Research next-generation encryption methods for data protection.
        \item Develop improved techniques for anonymization and pseudonymization that balance privacy with data utility.
        \item Investigate the potential of homomorphic encryption in enabling secure data analysis on encrypted data.
    \end{itemize}
    \item \textbf{Explainable AI (XAI) for HDI}:
    \begin{itemize}
        \item Advance methods for making complex AI decision-making processes interpretable to non-expert users.
        \item Develop user-friendly interfaces for explaining algorithmic decisions.
        \item Investigate the impact of XAI on user trust and engagement with data systems.
    \end{itemize}
    \item \textbf{Cross-Cultural HDI Implementation}:
    \begin{itemize}
        \item Conduct comparative studies on HDI implementation across different cultural contexts.
        \item Develop culturally adaptive HDI frameworks that respect local norms while upholding universal ethical principles.
        \item Investigate the impact of cultural differences on user perceptions of data privacy and agency.
    \end{itemize}
    \item \textbf{Quantifying HDI Impact}:
    \begin{itemize}
        \item Develop metrics and methodologies for measuring the effectiveness of HDI implementations.
        \item Conduct longitudinal studies on the impact of HDI on organizational performance and user trust.
        \item Investigate the economic implications of HDI adoption for businesses and industries.
    \end{itemize}
    \item \textbf{HDI in Emerging Technologies}:
    \begin{itemize}
        \item Explore the application of HDI principles in emerging fields such as the Internet of Things (IoT), edge computing, and quantum computing.
        \item Investigate the implications of HDI for human-AI collaboration and coexistence.
        \item Develop HDI frameworks for immersive technologies like augmented and virtual reality.
    \end{itemize}
    \item \textbf{HDI and Data Ecosystems}:
    \begin{itemize}
        \item Research methods for implementing HDI principles in complex, interconnected data ecosystems.
        \item Investigate the challenges and opportunities of HDI in decentralized data systems, including blockchain technologies.
        \item Develop strategies for maintaining HDI principles in data sharing and collaborative data environments.
    \end{itemize}
    \item \textbf{Regulatory Frameworks and HDI}:
    \begin{itemize}
        \item Analyze the interplay between HDI principles and evolving data protection regulations.
        \item Develop policy recommendations for incorporating HDI principles into legal frameworks.
        \item Investigate the potential for HDI to serve as a bridge between technological innovation and regulatory compliance.
    \end{itemize}
    \item \textbf{HDI Education and Literacy}:
    \begin{itemize}
        \item Develop comprehensive educational programs to enhance data literacy among the general public.
        \item Create curricula for training future data professionals in HDI principles and practices.
        \item Investigate effective methods for cultivating an HDI-centric organizational culture.
    \end{itemize}
    \item \textbf{HDI in Crisis and Pandemic Scenarios}:
    \begin{itemize}
        \item Explore the application of HDI principles in emergency data sharing situations.
        \item Develop frameworks for balancing public health needs with individual data rights.
        \item Investigate the long-term implications of emergency data practices on public trust and HDI implementation.
    \end{itemize}
\end{enumerate}

In conclusion, the HDI Framework offers a robust foundation for ethical, transparent, and human-centric data practices. As we navigate the complexities of our data-driven future, continued research and development in these areas will be crucial. By advancing our understanding and implementation of HDI, we can work towards a future where technological progress and human values are not in conflict, but in harmony. The journey towards fully realizing HDI principles within organizational practices is complex and challenging, but the potential benefits—a more ethical, transparent, and equitable data ecosystem that empowers individuals and upholds fundamental human values—are profound and far-reaching.

\section{Additional Informationl}
\subsection{Funding}
This work has received partial support from the National Project granted by the Ministry of Science, Innovation and Universities, Spain, under Grant PID2022-140974OB-I00, and from the Regional Government (JCCM) and the European Regional Development Funds (ERDF) through the INTECRA Project under Grant SBPLY/21/180501/000056.

\subsection{Declaration of interests}
The authors state that they have no financial interests or personal relationships that could have influenced the work presented in this article.

\subsection {Author contributions}
All authors contributed to the preparation and analysis of the article, as well as the drafting of the manuscript.

\bibliographystyle{unsrt}
\bibliography{references}  






\end{document}